\documentclass[twocolumn, mathlines]{aastex631}
\usepackage{graphicx}
\usepackage{txfonts}
\usepackage{booktabs}
\usepackage[normalem]{ulem}
\usepackage{multirow}
\usepackage{hyperref}
\usepackage{orcidlink}

\DeclareRobustCommand{\ion}[2]{%
\relax\ifmmode
\ifx\testbx\f@series
{\mathbf{#1\,\mathsc{#2}}}\else
{\mathrm{#1\,\mathsc{#2}}}\fi
\else\textup{#1\,{\mdseries\textsc{#2}}}%
\fi}

\begin{document}

\title{Environment and Gas Fraction in Type-2 AGN versus Non-AGN Galaxies}

\correspondingauthor{Jong-Hak Woo}
\email{jonghakwoo@gmail.com}

\author[0000-0002-5641-8102]{Jyoti Yadav}
\email{yadavjyoti636@gmail.com}
\affiliation{Instituto de Astrofísica de Canarias, Vía Láctea s/n, E-38205 La Laguna, Spain}
\affiliation{Departamento de Astrofísica, Universidad de La Laguna, E-38206 La Laguna, Spain}
\affiliation{Astronomy Program, Department of Physics and Astronomy, Seoul National University, Seoul 08826, Republic of Korea}

\author[0000-0002-8055-5465]{Jong-Hak Woo}
\affiliation{Astronomy Program, Department of Physics and Astronomy, Seoul National University, Seoul 08826, Republic of Korea}

\author[0009-0002-9526-5820]{Ashraf Ayubinia}
\affiliation{Astronomy Program, Department of Physics and Astronomy, Seoul National University, Seoul 08826, Republic of Korea}

\begin{abstract}
We investigate the environmental parameters and gas fraction (${\rm f_{gas}}$ ) properties of type~2 AGN and non-AGN galaxies, utilizing a large sample of galaxies from SDSS DR7 with z $\le$ 0.3. We find that the environment affects type~2 AGN and non-AGN galaxies in similar ways and does not impact the strength of AGN-driven outflows. The ${\rm f_{gas}}$ of type~2 AGN and non-AGN host galaxies show no variation between group and isolated environments, suggesting that host galaxy gas content is largely independent of large-scale environment. We find that type~2 AGN host galaxies possess systematically lower ${\rm f_{gas}}$ than their non-AGN counterparts when matched in stellar mass and star formation rate (SFR). This suggests that AGN activity plays a significant role in regulating the molecular gas reservoir and, consequently, the star formation processes within galaxies. We find that Type~2 AGNs exhibiting strong outflows are associated with higher gas fractions, higher star-formation rates, and younger stellar populations than those with weak or no outflows. This may indicate either concurrent star formation in gas-rich systems hosting powerful outflows, or a time delay between AGN activity and its effect on star formation consistent with a delayed AGN feedback scenario.

\end{abstract}


\section{Introduction} \label{sec:intro}
Active galactic nuclei (AGN) are widely considered 
as a key regulator
of star formation, by heating or expelling gas and thereby suppressing star formation \citep{Binney1995MNRAS.276..663B, Silk1998A&A...331L...1S, Ciotti2001ApJ...551..131C, Croton2006MNRAS.365...11C, McNamara2007ARA&A..45..117M, Nesvadba2008A&A...491..407N,  Cattaneo2009Natur.460..213C, Fabian2012ARA&A..50..455F, Heckman2014ARA&A..52..589H, Zinger2020MNRAS.499..768Z, Hopkins2024OJAp....7E..18H, Scharre2024MNRAS.534..361S}. 
However,  AGN-driven outflows can also trigger localized star formation through gas compression \citep{Zubovas2013MNRAS.433.3079Z, Ishibashi2014MNRAS.441.1474I, Schaye2015MNRAS.446..521S, Cresci2015A&A...582A..63C, Shin2019ApJ...881..147S, Nesvadba2020A&A...639L..13N, Schutte2022Natur.601..329S}. Multi-phase gas outflows have been observed in nearby AGNs \citep{Cicone2014A&A...562A..21C, Feruglio2015A&A...583A..99F, Harrison2018NatAs...2..198H, Garciabernete2021A&A...645A..21G, Almeida2022A&A...658A.155R, Ulivi2024A&A...685A.122U}. These outflows are potential evidence of energy injection from quasar-mode AGN feedback \citep{DiMatteo2005Natur.433..604D, Hopkins2008ApJS..175..356H, Garcia2021A&A...652A..98G, Navarro2021Natur.594..187M, RamosAlmeida2022A&A...658A.155R}. 
In nearby AGN, outflows have been directly imaged both on nuclear scales within dusty molecular tori and their immediate surroundings (tens of parsecs), as well as on the larger scales of circumnuclear disks (hundreds of parsecs) \citep{garcia2014A&A...567A.125G, Garcia2019A&A...632A..61G, Morganti2015A&A...580A...1M, Gallimore2016ApJ...829L...7G, Aalto2017A&A...608A..22A, Alonso2019A&A...628A..65A, Impellizzeri2019ApJ...884L..28I, Dominguez2020A&A...643A.127D}. Albeit with these observations, there is no clear consensus on whether these AGN-driven outflows have a significant impact on the cold gas content of their host galaxies \citep{Ho2008ApJ...681..128H, Cresci2015ApJ...799...82C, Russell2019MNRAS.490.3025R}.

In the AGN feedback scenario the star formation rates (SFR) and gas reservoirs in their host galaxies are expected to be significantly affected. The impact of AGN can be primarily observed through high-velocity outflows of ionized gas \citep{Karouzos2016ApJ...819..148K, Morganti2017FrASS...4...42M, Davies2020MNRAS.498.4150D}. Previous studies by \citet{Woo2017ApJ...839..120W, Woo2020ApJ...901...66W} showed that AGNs with strong outflows are linked to high specific SFRs (sSFR) and high Eddington ratio, whereas those with weak or no outflows exhibit lower SFRs and Eddington ratio . This suggests that AGN feedback may take time to impact galactic scales, a process known as delayed AGN feedback \citep[e.g.,][]{Cresci2018NatAs...2..179C}. \citet{Jarvis2020MNRAS.498.1560J} reported that the molecular gas reservoirs in the most powerful AGN are comparable to those in star-forming galaxies, representing a gas-rich phase of galaxy evolution with simultaneously high star formation and nuclear activity.

In this context it is essential to constrain gas content in AGN host galaxies for understanding the connection between supermassive black holes and their host galaxies. Cold molecular gas is commonly traced using CO \citep{Young1991ARA&A..29..581Y}. However, measuring the gas content based on CO is expensive for a large, representative galaxy sample. Thus, previous studies have relied on small, diverse samples of AGN host galaxies with cold gas measurements \citep{Cicone2014A&A...562A..21C, King2015ARA&A..53..115K, Bischetti2019A&A...628A.118B, Lutz2020A&A...633A.134L, Veilleux2020A&ARv..28....2V}. Alternatively,  molecular gas content can be constrained using dust emission in the thermal infrared \citep{Draine2007ApJ...663..866D, Scoville2014ApJ...783...84S} or dust attenuation from optical hydrogen Balmer lines \citep{Brinchmann2013MNRAS.432.2112B, Concas2019MNRAS.486L..91C}. Previous studies used optical nebular dust extinction and metallicity to estimate molecular gas mass (${\rm M_{H_{2}}}$) \citep{Yesuf2019ApJ...884..177Y}. For example, \citet{Saintonge2011MNRAS.415...32S} demonstrated that molecular gas is most strongly correlated with the NUV-r color, which is a tracer of the sSFR.

Environmental processes can significantly impact the gas reservoir in galaxies, potentially linking these mechanisms to the fueling of AGN activity. Previous studies explored the environments of AGN to investigate potential connections between AGN activity and environmental factors \citep{Miller2003ApJ...597..142M, Kauffmann2004MNRAS.353..713K, Gilmour2007MNRAS.380.1467G, Bradshaw2011MNRAS.415.2626B, Ellison2011MNRAS.418.2043E, Manzer2014ApJ...788..140M,Malavasi2015A&A...576A.101M, Koulouridis2018A&A...620A..20K, Steffen2023ApJ...942..107S, Barrows2023ApJ...951...92B, Li2023ApJ...944..168L, Erstegui2025A&A...699A.330E}.

Galaxy environment encompasses two key aspects: (a) the large-scale environment, which regulates the gas supply to galaxies, and (b) local-scale interactions, such as tidal forces between companion galaxies. These environmental influences can have opposing effects on AGN activity. For instance, AGN activity tends to decrease in high-density regions \citep{Carter2001ApJ...559..606C, Miller2003ApJ...597..142M}, and increases in field and lower density regions \citep{Kauffmann2004MNRAS.353..713K, Silverman2009ApJ...695..171S, Sabater2013MNRAS.430..638S, Lopes2017MNRAS.472..409L}. The AGN activity also increases in systems experiencing strong tidal interactions \citep{Petrosian1982Afz....18..548P, Koulouridis2006ApJ...639...37K, Ellison2011MNRAS.418.2043E, Duplancic2021MNRAS.504.4389D, Schechter2025ApJ...989..149S, Ellison2025OJAp....8E..12E}. Thus a systematic investigation of large-scale and local environmental effects, alongside internal ${\rm f_{gas}}$ properties, using a large and diverse galaxy sample is required to provide a clear understanding of how external and internal factors shape galaxy evolution.

In this study, we investigate the environmental effects and ${\rm f_{gas}}$ of type 2 AGN host galaxies compared to non-AGN galaxies. Since AGN galaxies overlap with the star-forming galaxy population, we classify the sample based on the presence of AGN to compare the environments and ${\rm f_{gas}}$ between AGN and non-AGN galaxies.
We aim to understand how external factors, such as environment, and intrinsic properties, such as gas content, differ among galaxy populations and their potential associations with evolutionary paths. 
This paper is organized as follows. In Section~\ref{sec:data}, we describe the data and the sample selection. We present the properties of galaxies in Section~\ref{sec:properties}. In Section~\ref{sec:analyis}, we present the analysis of P, F, and T parameters and ${\rm M_{H_{2}}}$ in galaxies. we present the results and discussion in Section~\ref{sec:results} and Section~\ref{sec:discussion}. We summarize our results in Section~\ref{sec:summary}.

\section{Sample Selection} \label{sec:data}
We use type~2 AGNs, of which the accretion disk and broad-line region are obscured by dusty material along the line of sight. Thus, type 2 AGNs allow us to reliably measure key properties, i.e., stellar mass and SFR of their host galaxies while reliable diagnostics using the emission from the narrow-line region can be applied to identify and characterize AGN activity. Thus, type~2 AGNs are well-suited for studying the co-evolution of galaxies based on a fair and consistent comparison with non-AGN galaxies.

\begin{figure}
    \includegraphics[width=0.98\linewidth]{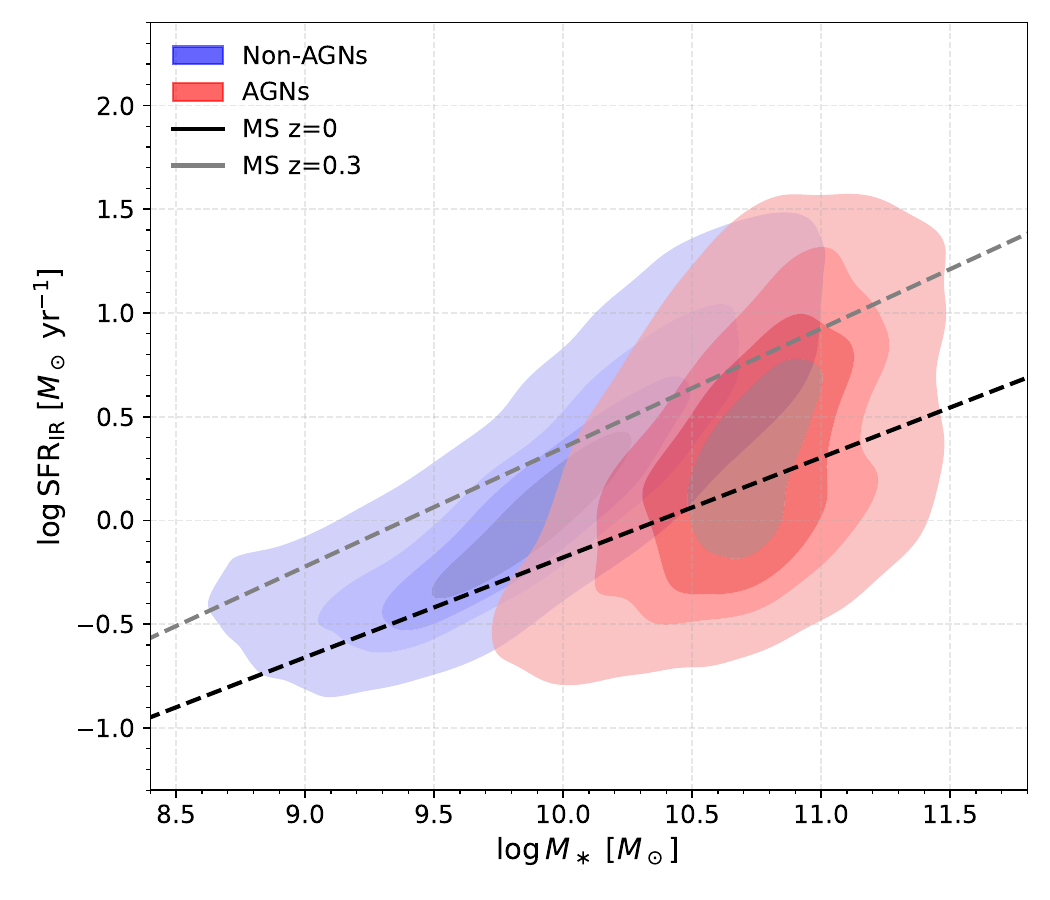}
    \caption{The distribution of type~2 AGN and non-AGN galaxies on star formation main sequence (SFMS). The sample spans a redshift range of 0--0.3. The solid and dashed line show SFMS at redshift 0 and 0.3 from \citet{Speagle2014ApJS..214...15S}.}
    \label{fig:SFMS_galaxies}
\end{figure}

We utilized a large sample of type~2 AGNs and non-AGN galaxies at z $\le$0.3 along with the SFR and ${\rm f_{gas}}$ from \citet{Woo2016ApJ...817..108W, Woo2017ApJ...839..120W}. The distribution of stellar mass and SFR is presented in Figure~\ref{fig:SFMS_galaxies}, respectively for AGNs and non-AGN galaxies.
The data from \citet{Woo2016ApJ...817..108W, Woo2017ApJ...839..120W} used in the study is drawn from the Max Planck Institute for Astrophysics and Johns Hopkins University (MPA/JHU) Catalog based on SDSS Data Release 7 \citep{Abazajian2009ApJS..182..543A}. The MPA/JHU SDSS catalog was cross-matched with Galaxy Evolution Explorer (GALEX) \citep{Bianchi2014AdSpR..53..900B} to obtain the NUV magnitudes of the galaxies, which were used to estimate the ${\rm M_{H_{2}}}$.
The MPA/JHU SDSS catalog also includes D$_n$4000 measurements from \citet{Brinchmann2004MNRAS.351.1151B}, which we have utilized to compare the ages of stellar populations in different outflow AGNs.
From the initial dataset of 235,922 galaxies at z$\leq0.3$, 112,726 emission-line galaxies are selected with a criterion requiring well-defined emission line profiles with an amplitude-to-noise (A/N) ratio greater than five for both the [O III] and H$\alpha$ lines. Then this sample is classified into three categories: AGNs (22752), composite systems (20709), and star-forming galaxies (69265) \citep{Woo2017ApJ...839..120W}.

Using the same sample, \citet{Woo2016ApJ...817..108W} presented a comprehensive study of the gas kinematics of type~2 AGN outflows 
and \citet{Woo2017ApJ...839..120W} reported the detailed comparison between SFR and AGN outflows. In this study we adopt [\ion{O}{iii}] outflow kinematics and SFRs from \citet{Woo2016ApJ...817..108W, Woo2017ApJ...839..120W} and \citet{Woo2017ApJ...839..120W} and provide a brief overview of the measurement procedure in Section~\ref{sec:properties}.

\subsection{Properties of Galaxies}\label{sec:properties}
\subsubsection{ \rm[\ion{O}{iii}] Kinematics}
\citet{Woo2016ApJ...817..108W} provided a full description of the sample. They derived gas kinematics by utilizing the [\ion{O}{iii}] emission line at, 5007\AA, which also traces the ionized gas outflows. 
The measurement of the flux-weighted center (first moment) and velocity dispersion (second moment) of the total line profile was derived from the best-fit model of the [\ion{O}{iii}] emission line.

The velocity dispersions were corrected for the SDSS instrumental resolution (55--60 km s$^{-1}$ at the [\ion{O}{iii}]). Objects with unresolved [\ion{O}{iii}] lines, low dispersions, or large fractional errors ($>1$) were excluded.

The strength of type~2 AGN outflow was estimated by dividing the stellar velocity dispersion by the gas velocity dispersion. The velocity and velocity dispersion of [\ion{O}{III}] were added in quadrature and denoted by ${\rm \sigma_{[O\,III]}^{'}}$. This outflow velocity dispersion ($\sigma_{[\rm O\,III]}^{'}$) was then normalized by the stellar velocity dispersion (${\rm \sigma_{*}}$) to remove the effect of gravitational kinematics. AGNs were classified as having strong (${\rm log(\frac{\sigma_{[O\,III]}^{'}}{\sigma_{*}})~\ge~0.3}$), weak (0~$\le$~${\rm log(\frac{\sigma_{[O\,III]}^{'}}{\sigma_{*}}}$)~$<$~0.3), and no outflow (${\rm log(\frac{\sigma_{[O\,III]}^{'}}{\sigma_{*}}}$)~$<$~0). We estimated the bolometric luminosity as ${\rm L_{bol}} = {\rm L_{[O\,III]}} \times 3500$, using the extinction-uncorrected [\ion{O}{III}] luminosity \citep{Heckman2004ApJ...613..109H}. The ${\rm L_{bol}}$ of AGN varies from $10^{42}$ to $10^{45.5}$ erg s$^{-1}$ for our sample. In addition to the ionized gas kinematics, the catalog also contains the Eddington AGN luminosity, Eddington ratio, and black hole mass of each object.

\subsubsection{Star Formation Rate}
The SFR of the galaxy used in this study was taken from \citet{Woo2017ApJ...839..120W} and \citet{Ellison2016MNRAS.455..370E}, who used infrared luminosity as a proxy for star formation activity. Note that the IR emission is primarily driven by the absorption of ultraviolet and visible light by dust. Thus, The far-IR emission is closely related to the amount of dust-obscured star formation, providing a reasonble estimate of SFR in cases where AGN contaminates other SFR indicators, i.e., the H$\alpha$ emission line and the ultraviolet continuum.

Far-IR luminosity was adopted from \citet{Ellison2016MNRAS.455..370E}, who determined the IR luminosities of $\sim$330,000 SDSS galaxies, using artificial neural network methods. They adopt the mean ${\rm L_{IR}}$ value of the 20 best trained networks and assign the associated error (${\rm \sigma_{ANN}}$) as the scatter amongst the network output. These techniques were developed and refined through testing and training on a matched sample of 1,136 galaxies from the Herschel-SDSS Stripe 82 dataset. The SFR was then estimated by converting the IR luminosity using the conversion formula provided by \citet{Kennicutt1998ARA&A..36..189K}.

\begin{equation}
\rm{\log L_{IR}\,(erg\,s^{-1})} = \rm{\log SFR\,(M_{\odot}\,yr^{-1}) + 43.591}
\end{equation}

The method proposed by \citet{Ellison2016MNRAS.455..370E} has been demonstrated to provide reliable SFR estimates for type~2 AGN host galaxies. \citealt{Ellison2016MNRAS.455..370E} recommended using a threshold of ${\rm \sigma_{ANN}} < 0.1$; however, this criterion resulted in the exclusion of a large number of AGN. To address this limitation and increase the sample size,  \citealt{Woo2017ApJ...839..120W} adopted a more relaxed criterion of ${\rm \sigma_{ANN}} < 0.3$ after comparing with the directly measured IR luminosity of a subsample. Following this approach, we used galaxies with ${\rm \sigma_{ANN}} < 0.3$ for our analysis. 
We present the results based on SFR estimates derived from IR luminosity. For a consistency check, we compare SFRs determined by different methods in Appendix~\ref{sec:comparison_sfr}.

\subsection{Molecular Gas Mass Estimation}
Several studies have developed empirical scaling relations that enable statistical predictions of ${\rm f_{gas}}$ based on NUV-r color, particularly using large survey data sets such as GALEX and SDSS. Studies have shown that NUV-r color is closely correlated with the gas content of galaxies \citep{Catinella2010MNRAS.403..683C, Catinella2012A&A...544A..65C, Saintonge2011MNRAS.415...32S}. Galaxies with bluer NUV-r colors typically exhibit higher star formation activity, which is generally associated with larger gas reservoirs \citep{Salim2014SerAJ.189....1S}. These correlation supports the use of NUV-r color as a reliable indirect tracer of the ${\rm M_{H_{2}}}$ in the galaxy. We used the scaling relation by \citet{Saintonge2011MNRAS.415...32S} to estimate the ${\rm M_{H_{2}}}$. Since our targets are type~2 AGNs, we expect the colors to remain largely unaffected by the AGN continuum.

\begin{equation}
    \rm{log(M_{H_{2}}/M_{*})} = \rm{-0.293 ( NUV - r - 3.5 ) - 1.349}
\end{equation}
The NUV, and r magnitudes were taken from GALEX and SDSS. We use the SDSS aperture-corrected model r magnitude (modelMag\_r), which are intended to represent the total galaxy flux, while the GALEX NUV magnitude are based on Kron photometry and similarly approximate the integrated emission from the entire galaxy. Both measurements therefore trace global photometry, and we do not expect them to introduce a significant systematic bias in the NUV--r color. The NUV-r colors have been corrected for Galactic extinction following the methodology of \citealt{Wyder2007ApJS..173..293W}, who adopted extinction coefficients of A($\lambda$)/E(B-V) = 2.751 for the SDSS r band and A($\lambda$)/E(B-V) = 8.2 for the GALEX NUV band. Based on these values, the applied correction to the NUV-r colors is ${\rm A_{NUV}-A_{r} = 1.9807\times A_{r}}$, where the extinction in the r band (${\rm A_{r}}$) is obtained from the SDSS database. 
We estimated the ${\rm f_{gas}}$ as follows
\begin{equation}
\rm{f_{gas}} = \rm{M_{{H_{2}}}/(M_{\ast}+M_{{H_{2}}}) }
\end{equation}

The stellar masses of galaxies were taken from the MPA/JHU Catalogue. A subset of galaxies exhibit infrared luminosities exceeding 10$^{11.5}$ L${\odot}$, placing them in the regime of luminous infrared galaxies (LIRGs). To avoid biases, we excluded these systems for the ${\rm f_{gas}}$ analysis.

We estimate the depletion time as:
\begin{equation}
    \rm{t_{dep}} = \rm{M_{{H_{2}}}/SFR}
\end{equation}

\citealt{Saintonge2011MNRAS.415...32S} used a sample of COLD GASS galaxies with CO(1--0) measurements from the IRAM 30m telescope. Several galaxies in our sample also have observed H$_2$ masses available from the xCOLD GASS survey. We compared our estimated H$_2$ masses with these observed values and found that the associated uncertainty in the derived ${\rm f_{gas}}$ is approximately 0.15 dex, indicating a relatively low level of scatter and suggesting that the estimates are robust for comparative and statistical analyses. The same relation was also employed by \citet{Luo2019ApJ...874...99L} to study molecular ${\rm f_{gas}}$ in AGN host galaxies, where they reported a systematic uncertainty of up to a factor of 2 in the total ${\rm f_{gas}}$ (HI + H$_2$). While the photometric method is subject to substantial systematic uncertainties, it remains a useful tool for statistical studies of large galaxy samples. 

NUV magnitudes from GALEX were missing for some galaxies in the sample, limiting our ability to estimate ${\rm M_{H_{2}}}$ for the entire dataset. We reliably estimated ${\rm M_{H_2}}$ and ${\rm f_{gas}}$ for 10757 galaxies which have available NUV and r colors and ${\rm L_{IR}}$ $<$ 10$^{11.5}$ L$_{\odot}$, comprising 2446 type~2 AGNs and 8311 non-AGNs.

\section{Analysis} \label{sec:analyis}
\subsection{P, F and T Parameter Estimation}
We investigated the influence of galaxy environments on type~2 AGN and non-AGN galaxies. We utilized the galaxy group catalog compiled by \citet{Yang2007ApJ...671..153Y}, which includes galaxies from the Main Galaxy Sample of the NYU Value-Added Galaxy Catalogue (NYU-VAGC) within the redshift range 0.01 $\le$ z $\le$ 0.20. \citealt{Yang2007ApJ...671..153Y} employed a modified version of the halo-based group finder algorithm to classify the galaxies into groups. The group finder is highly effective at grouping galaxies within common dark matter halos. It also performs consistently well for low-mass systems, accurately identifying isolated galaxies in small dark matter halos. \citet{Yang2007ApJ...671..153Y} classified galaxies as group members if they are linked to other galaxies within a projected separation of 0.05 times the mean inter-galaxy separation and a line-of-sight redshift separation of 0.3 times the mean inter-galaxy separation at the corresponding redshift. Galaxies that do not have any neighbors within these linking lengths are identified as isolated/single galaxies.

To study the effect of the environment on the galaxies, we defined three different environment factors (P, F, and T parameters) as 

\begin{equation}
    \rm{P}= \rm{\sum_{i=1}^{N} \frac{M_{i}}{r_{i}} \qquad}
    \rm{F}= \rm{\sum_{i=1}^{N} \frac{M_{i}}{r_{i}^2} \qquad}
    \rm{T}= \rm{\sum_{i=1}^{N} \frac{M_{i}}{r_{i}^3}}
\end{equation}

where ${\rm M_{i}}$ is stellar mass in (${\rm M_{\odot}}$) and ${\rm r_{ i}}$ is the projected distance (in Mpc) of i'th nearest neighbor and N+1 is the number of galaxies in the group. 

P denotes the total gravitational potential on a galaxy by other members of its group. However, the quantities F and T, as defined in this analysis, do not correspond directly to physical vector quantities such as net gravitational force or tidal field strength. F and T represent scalar sums of the magnitudes of gravitational forces and tidal field strengths, respectively, contributed by all other group members. Since both gravitational force and tidal field strength are inherently vector, this scalar treatment does not capture the true net vector quantities experienced by a galaxy. See Appendix~\ref{sec:pft_analysis} for examples demonstrating the use of PFT.
 
Cross-matching the Yang catalog with our initial galaxy sample (112726) resulted in a total of 88498 galaxies, comprising Type~2 AGN, and non-AGN galaxies. We classified galaxies as group members if they were part of systems containing two or more galaxies. In contrast, galaxies that were the only identified members of their group were considered isolated or single galaxies. Based on this classification, our dataset consists of 24430 (AGNs and non AGNs) galaxies located in groups (with group membership of two or more) and 64068 (AGNs and non AGNs) galaxies identified as isolated systems. We utilized these classifications to examine and compare the properties of group and isolated galaxies in Section~\ref{sec:results}. We have 7913 AGNs, and 16517 non-AGNs in group and 14001 single AGNs and 50067 non-AGN host isolated galaxies.

\begin{figure}
    \includegraphics[width=0.98\linewidth]{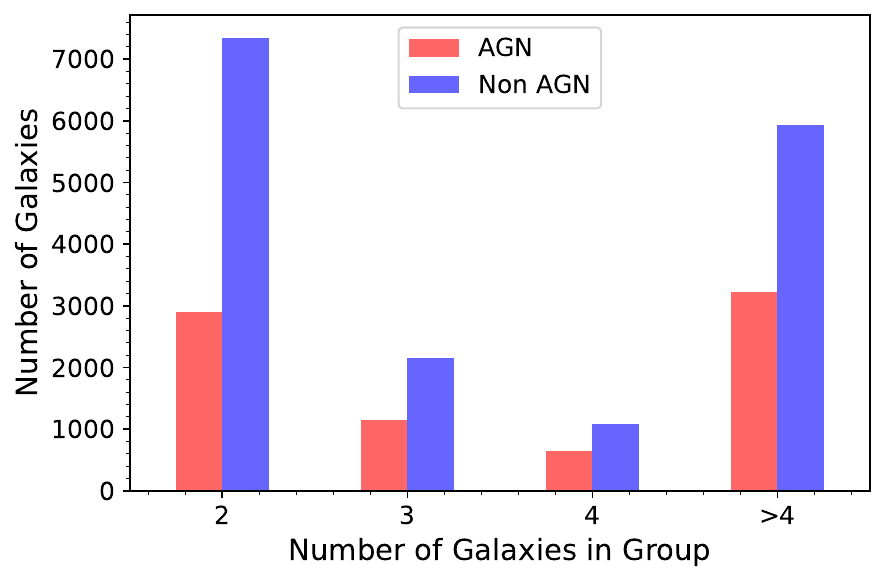}
    \caption{The distribution of type~2 AGN and non-AGN galaxy counts across groups with varying numbers of member galaxies.}
    \label{fig:count_galaxies}
\end{figure}

 Figure~\ref{fig:count_galaxies} presents the counts of AGN and non-AGN galaxies across groups of varying sizes.

\section{Results} \label{sec:results}
To investigate the influence of environment and gas content on type~2 AGN activity, we first examined the role of large-scale environmental effects by estimating the P, F, and T parameters and comparing their distributions between type~2 AGN and non-AGN galaxies, as described in Section~\ref{sec:large_scale_environment}. We then assessed local environmental influences by calculating the same parameters based on the nearest neighbor within a projected distance of 150 kpc (Section~\ref{sec:local_scale_environment}). We also estimated the type~2 AGN fraction as a function of the P, F, and T parameters in Section~\ref{sec:AGN_fraction}.

\citet{Ellison2019MNRAS.482.5694E} emphasize that matching control samples on parameters beyond stellar mass is crucial for a robust and unbiased comparison. To examine the effects of environment on galaxy properties in more detail, we analyzed the gas fraction (${\rm f_{gas}}$) and ${\rm M_{H_{2}}}$ for subsamples matched in stellar mass and star formation rate, as described in Section~\ref{sec:group_vs_single}.  We further divided both the type~2 AGN and non-AGN samples into group and isolated galaxies. This approach allowed us to directly compare group type~2 AGNs with group non-AGNs, and isolated type~2 AGNs with isolated non-AGNs. We also compared group versus isolated type~2 AGNs and group versus isolated non-AGNs to specifically evaluate how the environment influences each type of galaxy.

To explore the connection between gas content and AGN activity, we compared the ${\rm f_{gas}}$ and ${\rm M_{H_{2}}}$ across the full type~2 AGN and non-AGN galaxy samples, as presented in Section~\ref{sec:gas_fraction_comparison}.

\begin{figure*}
    \centering
    \includegraphics[width=0.75\linewidth]{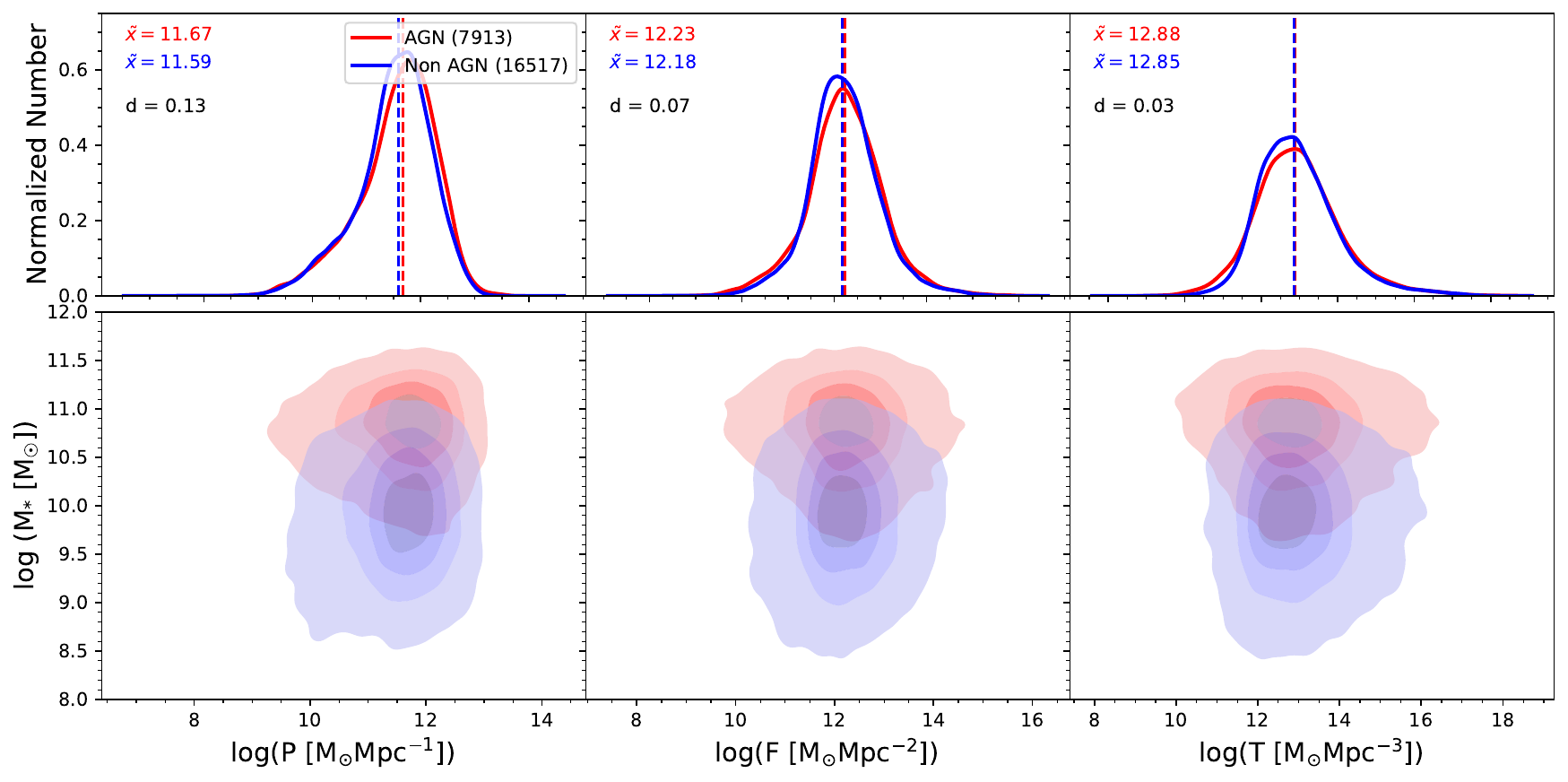} \\
  \caption{The P, F, and T parameter distribution of galaxies. The top panel shows the histogram of P, F, and T parameters for type~2 AGN (red) and non-AGN galaxies (blue) respectively. The dashed line shows the median value of each distribution. d represents the Cohen's d value which is discussed in Section~\ref{sec:large_scale_environment}. }
    \label{fig:AGN_SFG_pft}
\end{figure*}

\begin{figure*}
    \centering
     \includegraphics[width=0.75\linewidth]{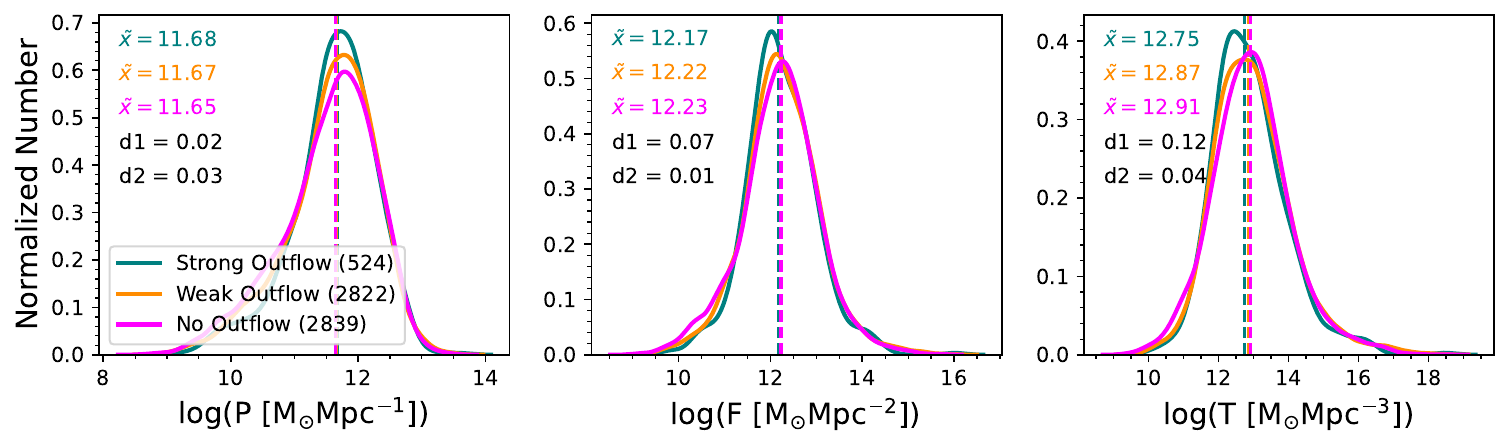}  \\
 \includegraphics[width=0.75\linewidth]{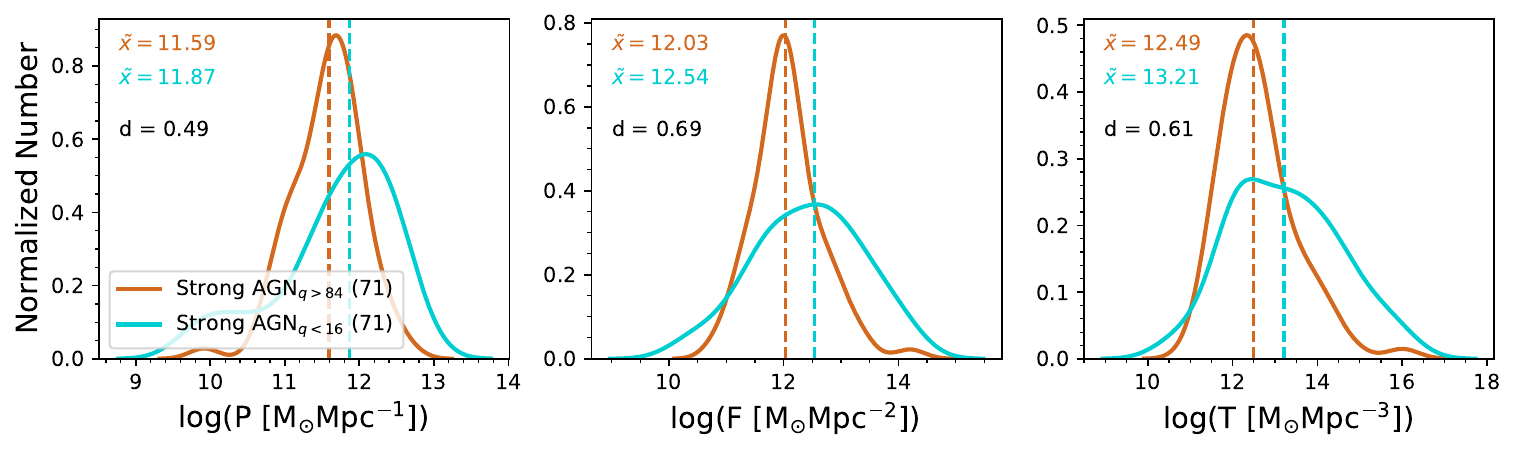}
    \caption{The top panel shows the histogram of P, F, and T parameters for strong (teal), weak (orange), and no outflow AGNs (magenta) respectively. The bottom panel shows the histogram of P, F, and T parameters for Strong outflow having sSFR above 84 (brown) and below 16 percentiles (cyan) respectively. The d1 and d2 are Cohen's d value which are discussed in Section~\ref{sec:large_scale_environment}.}
    \label{fig:PFT_swn_16_84}
\end{figure*}

\subsection{Effect of Large Scale Environment} \label{sec:large_scale_environment}
To explore how the environment influences star formation and AGN activity, we analyzed P, F, and T parameters and used Cohen’s d test \citep{Cohen1990CEUS...14Q..71.} to measure differences between the samples. Cohen's d is an effect size measure used to quantify the difference between two means relative to the pooled standard deviation. The effect size is defined as:
\begin{equation}
    \rm{d}= \rm{\frac{\bar{X}-\bar{Y}}{s_{p}}}
\end{equation}
where $\rm {\bar{X}}$ and ${\rm \bar{Y}}$ are sample means and ${\rm s_{p}}$ is the pooled standard deviation and is defined as follows:
\begin{equation}
    \rm{s_{p}}= \rm {\sqrt{\frac{(n-1)s_{x}^2 + (m-1)s_{y}^2}{n+m-2}}}
\end{equation}
where n and m are the two sample sizes and ${\rm s_{x}}$ and ${\rm s_{y}}$ are the sample variances respectively.

Cohen’s d assumes the data are normally distributed and is sensitive to outliers as it relies on the mean and standard deviation. Thus we used a robust method to calculate effect size by comparing medians instead of means and using the median absolute deviation (MAD) instead of standard deviation. This makes the results less affected by outliers and skewed data. We combined the differences within each group using a pooled MAD, similar to how pooled standard deviation is used in Cohen’s d. This method gives a more accurate and reliable effect size, especially when the data are not normally distributed.

We adopt this robust Cohen’s d for non-normal distributions to assess the separation between subsamples (e.g., type~2 AGN vs. non-AGN), as it captures the magnitude of difference, unlike other significance tests (e.g., Kolmogorov-Smirnov or Mann--Whitney U) that only indicate whether a difference exists. We use Cohen’s d value = [0.2--0.5), [0.5--0.8), and [0.8--1] to interpret observed effect sizes as small, medium, or large, respectively. The d values in each plot indicate the corresponding Cohen’s d measurement, reflecting the magnitude of the difference. The $\bar{x}$ in each plot, displayed in the same color as the corresponding distribution, indicates the median of that distribution.

We first compared the environmental parameters of type~2 AGN and non-AGN galaxies (Figure~\ref{fig:AGN_SFG_pft}, top panel). Type~2 AGN hosts (red) are, on average, more massive than non-AGN galaxies (blue), consistent with previous studies \citep{Woo2017ApJ...839..120W}. The Cohen’s d values for the P, F, and T parameters (0.13, 0.07, and 0.03, respectively) indicate minimal differences in environmental characteristics, suggesting that type~2 AGN hosts and non-AGN galaxies inhabit similar environments. 

AGN driven outflow strength varies widely, with some galaxies exhibiting powerful outflows and others showing little or no detectable activity. Moreover, the specific SFR (sSFR) has been found to correlate with outflow strength \citep{Woo2020ApJ...901...66W}. To test whether environment influences AGN outflow strength and sSFR, we compared the P, F, and T parameters for strong, weak, and no outflow AGNs (Figure~\ref{fig:PFT_swn_16_84}, top panel), excluding AGNs with unresolved [\ion{O}{iii}] lines, low velocity dispersions, or large fractional errors. The sample includes 524 strong, 2822 weak, and 2839 no outflow AGNs. Cohen’s d values, d1 (strong vs. weak) and d2 (weak vs. none), indicate that the distributions of P, F, and T are similar across these subgroups, suggesting that large-scale environment does not significantly affect outflow strength.

Strong outflow AGNs exhibit, on average, higher sSFR than weak or non-outflow AGNs \citep{Woo2020ApJ...901...66W}, though sSFR shows considerable scatter within this group. To investigate this variation, we divided strong outflow AGNs into high-sSFR (above the 84th percentile, q$>$84) and low-sSFR (below the 16th percentile, q$<$16) subsamples, each containing 71 galaxies. Comparing their P, F, and T parameters (Figure~\ref{fig:PFT_swn_16_84}, bottom panel) yields Cohen’s d values of 0.49, 0.69, and 0.61, indicating moderate environmental differences. This suggests that environment may influence sSFR in strong outflow AGNs, though statistical significance is limited by the small sample size.

\begin{figure*}
    \centering
    \includegraphics[width=0.85\linewidth]{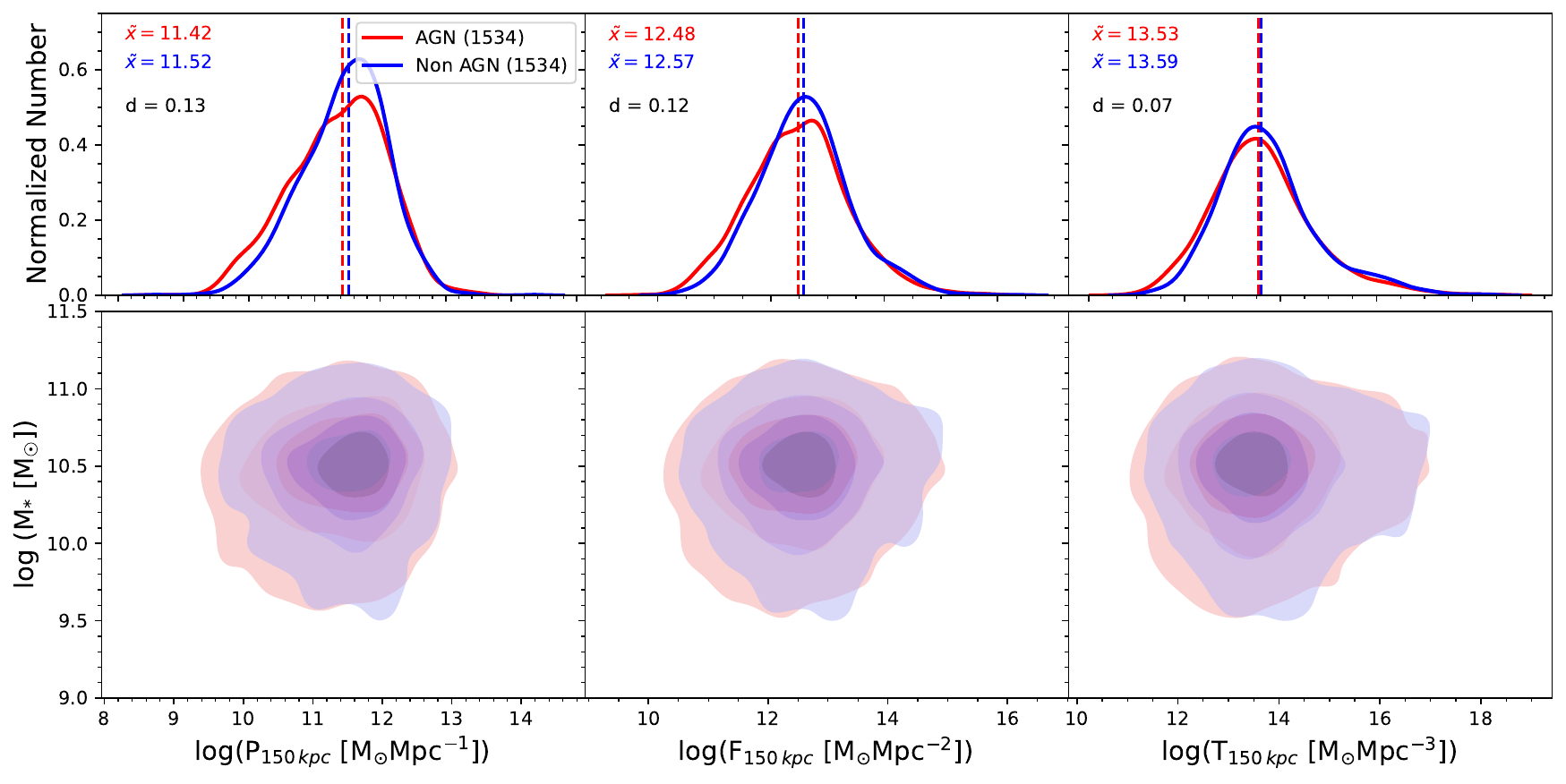} 
    \includegraphics[width=0.85\linewidth]{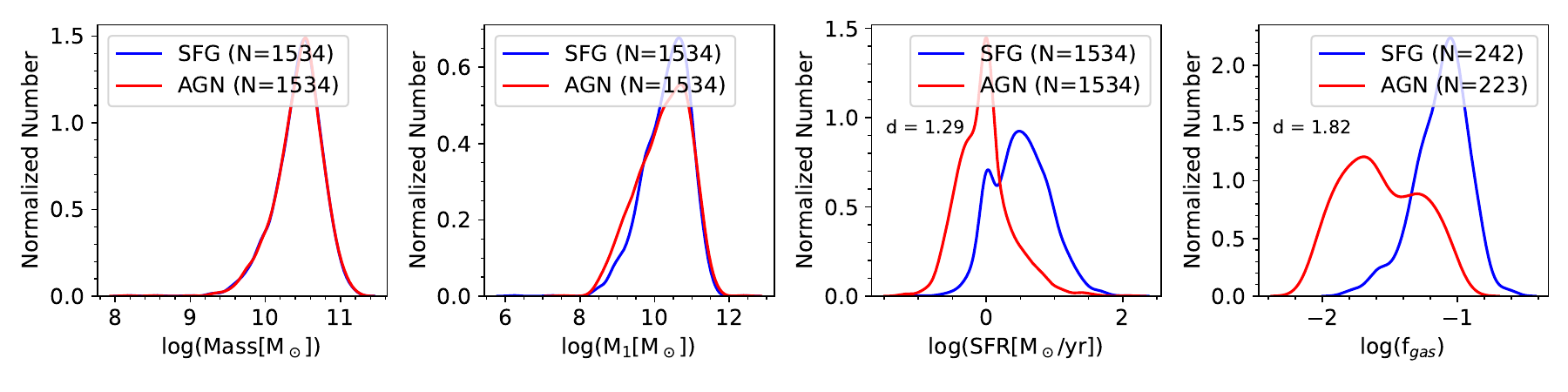}
    \caption{The P, F, and T parameter distribution of galaxies from only the first nearest neighbor within 150 kpc. The top panel shows the histogram of P, F, and T parameters for type~2 AGN (red) and non-AGN galaxies (blue) respectively. The middle panel shows the distribution of stellar mass. The bottom panel presents the properties of AGN and non-AGN host galaxies matched in stellar mass. The Comparison shows that AGN hosts exhibit systematically lower SFRs and ${\rm f_{gas}}$ compared to non-AGN hosts when controlled for M${*}$, and environments.}
    \label{fig:near_neighbour_effect}
\end{figure*}

\begin{figure*}
    \centering
    \includegraphics[width=0.95\linewidth]{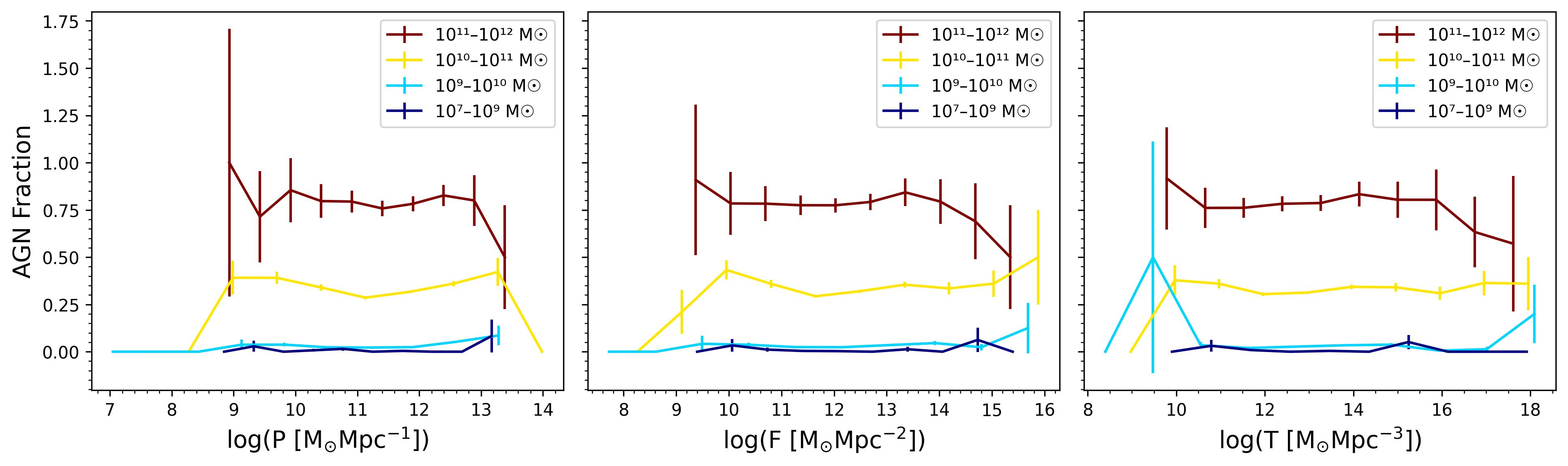}
    \caption{Type~2 AGN fraction as a function of P, F, and T environment parameters.}
    \label{fig:agn_fraction}
\end{figure*}

\subsection{Effect of Local Environment -- in Terms of the Nearest Neighbour}\label{sec:local_scale_environment}
The nearest neighbor can exert a substantial influence on a host galaxy, with tidal forces enhancing star formation or even triggering AGN activity \citep{Hernquist1989Natur.340..687H}. Such interactions highlight the critical role of close galactic neighbors in shaping galaxy evolution \citep{Arp1966apg..book.....A, Sanders1996ARA&A..34..749S, Struck1999PhR...321....1S, Saintonge2012ApJ...758...73S}, and \citet{Patton2013MNRAS.433L..59P} show that galaxies up to 150 kpc away can influence host properties such as SFR.

To examine the effect of the nearest neighbor on type~2 AGN and non-AGN galaxies, we considered neighbors within 150 kpc and matched the samples in stellar mass. Both samples were divided into 50 stellar mass bins, with galaxies in the same bin considered comparable. To avoid bias from unequal bin sizes, we randomly matched the number of galaxies in each bin, resulting in two final subsamples of 1534 galaxies each.

The distributions of ${\rm P_{150kpc}}$, ${\rm F_{150kpc}}$, and ${\rm T_{150kpc}}$ (Figure~\ref{fig:near_neighbour_effect}, top panel) show negligible differences between type~2 AGN and non-AGN galaxies, with Cohen’s d values of 0.13, 0.12, and 0.07, respectively. This indicates that the influence of the nearest neighbor is similar across both populations.

We compared the SFR and ${\rm f_{gas}}$ of matched AGN and non-AGN samples Figure~\ref{fig:near_neighbour_effect}, bottom panel), where M and ${\rm M_{1}}$ denotes the mass of main galaxy and nearest-neighbor mass. AGN host galaxies exhibit systematically lower SFR and ${\rm f_{gas}}$ than their non-AGN counterparts, despite having comparable stellar masses and local environments. The large Cohen’s d values of 1.29 for SFR and 1.82 for ${\rm f_{gas}}$ indicate a significant suppression of both star formation and gas content associated with AGN activity. The broader ${\rm f_{gas}}$ distribution among AGN hosts further suggests that feedback effects are not uniform: galaxies hosting long-lived or recurrent AGNs \citep{Harrison2024Galax..12...17H} may have undergone substantial gas depletion, whereas those with recently triggered AGN activity may not yet show measurable impacts on their gas reservoirs.

\subsection{Impact of Environment on type~2 AGN Fraction}\label{sec:AGN_fraction}
To examine the role of the environment in triggering AGN activity, we analyzed the type~2 AGN fraction as a function of the  P, F, and T parameters (Figure~\ref{fig:agn_fraction}). AGNs are more frequently found in massive galaxies. These massive galaxies can strongly influence their local environment. However, they may still show low P, F, and T values if they lack nearby galaxies of similar mass. This makes it challenging to separate the influence of the environment from the effects of stellar mass. If this mass dependence is not properly accounted for, it can introduce bias when analyzing the relationship between AGN fraction and environmental parameters. As a result, the interpretation of AGN fraction trends may be skewed. To account for this bias, we divided the galaxy sample into different stellar mass bins and investigated the type~2 AGN fraction with respect to P, F, and T parameters within stellar mass bins. We found that the type~2 AGN fraction does not vary with respect to these environmental parameters across all stellar mass bins. This suggests that environmental factors, quantified by P, F, and T, do not play a significant role in AGN activity. Instead, intrinsic properties of galaxies, such as stellar mass, and gas content, could be the primary drivers of AGN fueling mechanisms. We show the comparison of internal properties of the type~2 AGN and non-AGN galaxies in the next sections.

\begin{figure*}
    \centering
    \includegraphics[width=0.43\linewidth]{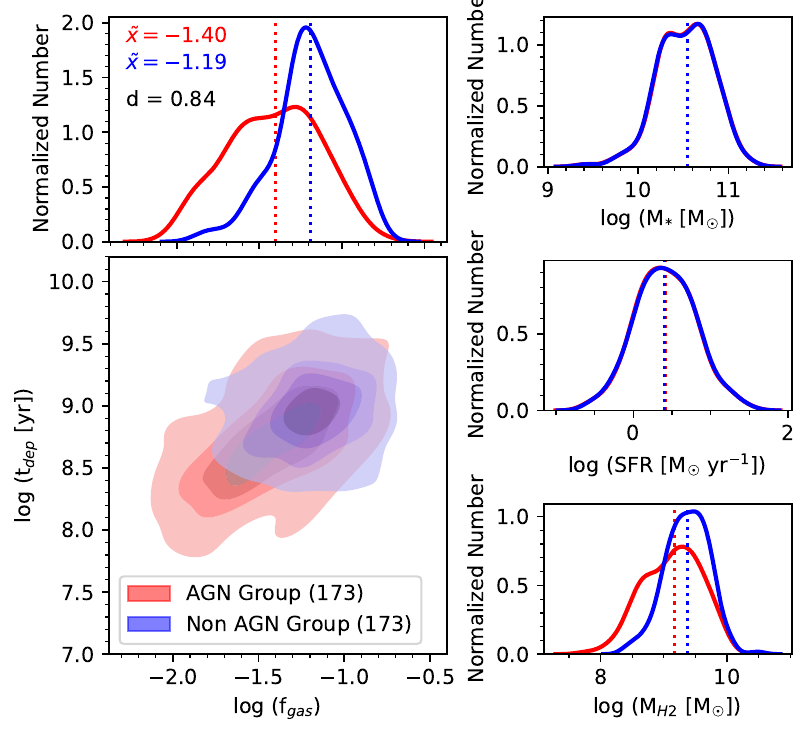} 
    \includegraphics[width=0.43\linewidth]{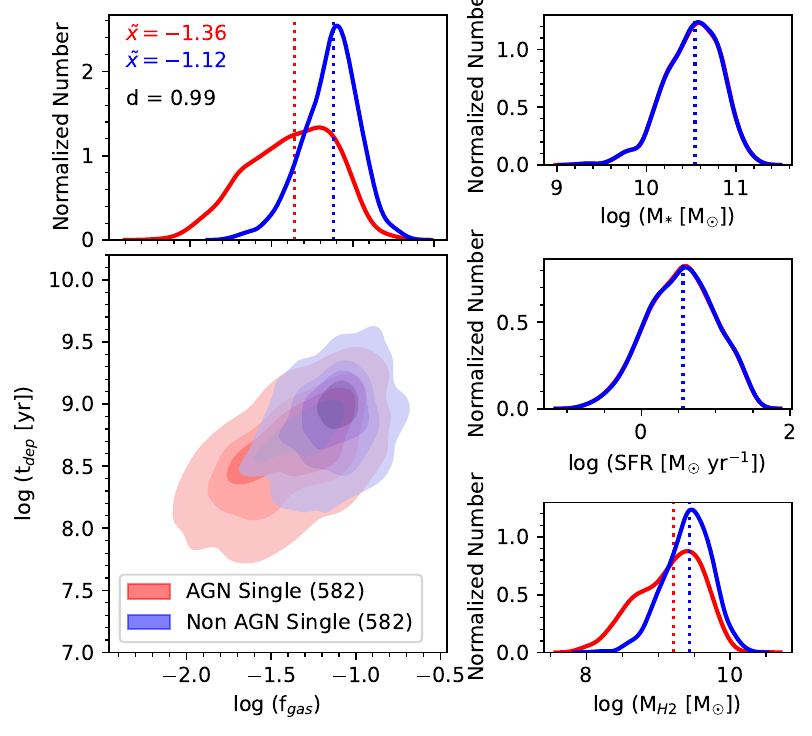}
    \caption{Comparison of properties of type~2 AGN (red) and non-AGN galaxies (blue) in group and isolated environment. The left panels show the comparison of type~2 AGN and non-AGN galaxies in a group, while the right panels show the comparison in an isolated environment. The dashed line shows the median value of each distribution.}
    \label{fig:match_AGN_SFG_group_single}
\end{figure*}

\begin{figure*}
    \centering
    \includegraphics[width=0.43\linewidth]{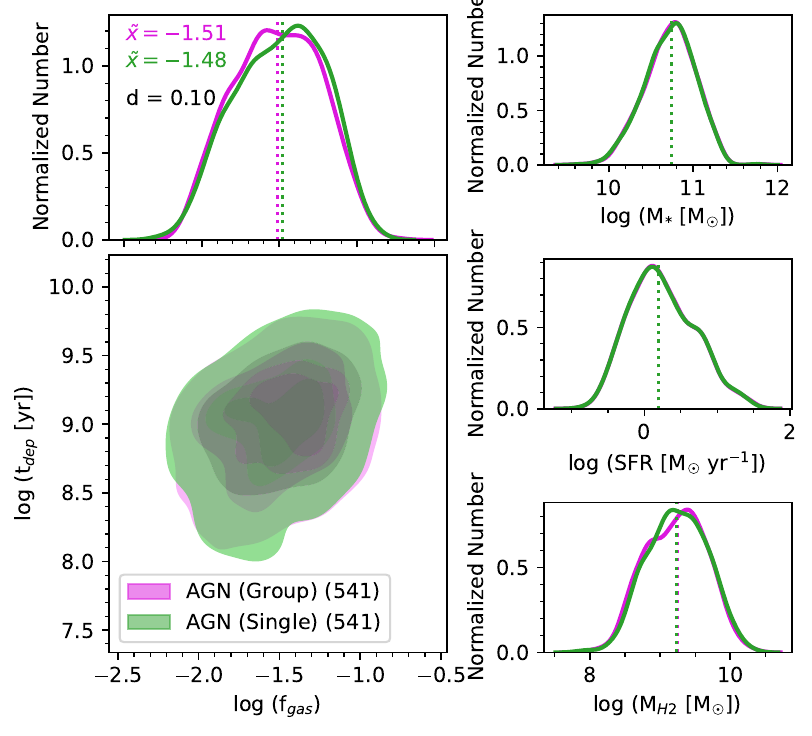} 
    \includegraphics[width=0.43\linewidth]{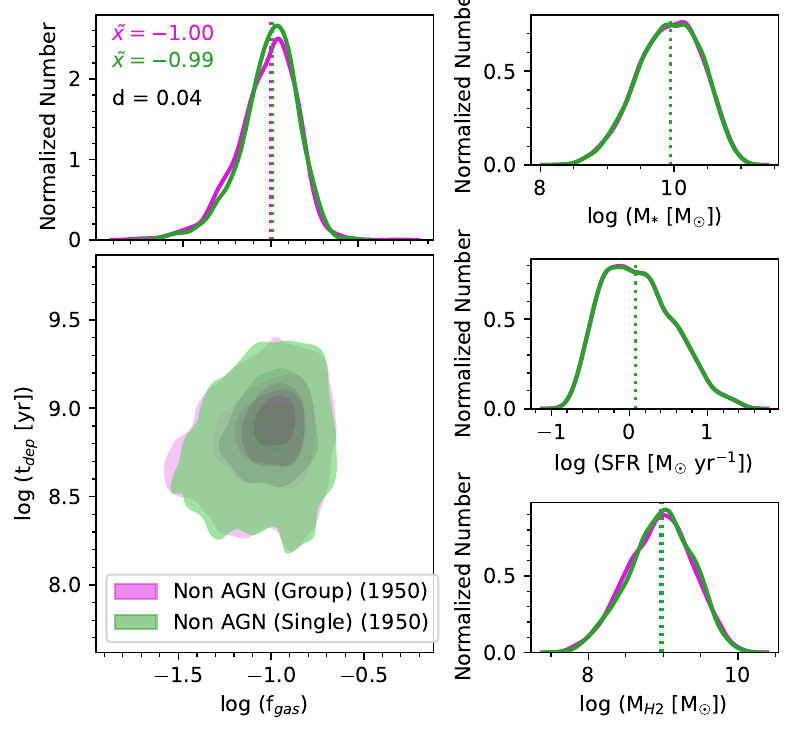}
    \caption{Comparison of galaxy properties in group and isolated environments. The left panels show the comparison of properties of type~2 AGN host galaxies in a group (purple) and isolated environments (green) respectively. The right panels show the comparison of the properties of non-AGN galaxies in group and isolated environments. The dashed line shows the median value of each distribution.}
    \label{fig:AGN_SFG_group_single}
\end{figure*}

\subsection{Comparison of Group vs. Isolated Environments and Intra-Group Variations}\label{sec:group_vs_single}
The role of environment in shaping galaxy properties and evolution remains an active area of research. External processes such as mergers \citep{Matzko2022MNRAS.514.4828M} and tidal interactions \citep{Martig2008MNRAS.385L..38M} can drive gas inflows, triggering star formation and AGN activity. However, several studies suggest that environment may not be the primary driver of AGN activity. The AGN fraction appears roughly constant across local densities \citep{Miller2003ApJ...597..142M, Martini2006ApJ...644..116M, Rodriguez2019MNRAS.486..344R}, and internal secular processes are proposed as the dominant mechanism for AGN triggering \citep{Man2019MNRAS.488...89M}.

To investigate environmental effects, we compared type~2 AGN and non-AGN galaxies in both group and isolated environments. Comparisons were made between group AGNs and group non-AGNs, and between isolated AGNs and isolated non-AGNs, enabling an assessment of AGN impact under similar environmental conditions. We further compared group versus isolated systems within each population. We divided the galaxies into bins of stellar mass and SFR, forming a two-dimensional grid. We randomly select an equal number of objects from the AGN and non-AGN samples, where the number is set by the smaller population in that bin. Bins that are populated by only one sample are excluded from the matched dataset. This procedure ensures that the two samples have equal sizes and similar joint distributions in stellar mass and SFR. Cross-matching in stellar mass and SFR resulted in 173 group galaxies and 582 isolated galaxies in each sample.

The Cohen’s d values indicate moderate to large differences in ${\rm f_{gas}}$ between type~2 AGNs and non-AGN galaxies, in both group (d=0.84) and isolated (d=0.99) environments. These results are consistent across different SFR estimates, including IR-derived SFRs, demonstrating that non-AGN galaxies maintain higher ${\rm f_{gas}}$ than AGN hosts at fixed stellar mass and SFR. This suggests that AGN activity may deplete the gas reservoir, regulating star formation and influencing host galaxy evolution.

We also examined differences in ${\rm f_{gas}}$ between group and isolated galaxies within each population. Cross-matching yielded 541 type~2 AGNs and 1950 non-AGNs across both environments. Type~2 AGN hosts exhibit comparable ${\rm f_{gas}}$ in group and isolated systems (Cohen’s d = 0.10), and non-AGN galaxies show a similarly negligible difference (d = 0.04). Results based on alternative SFR estimates remain consistent (Appendix~\ref{sec:comparison_sfr}). These findings suggest that environment exerts only a minor influence on ${\rm f_{gas}}$, implying that the molecular gas content is primarily regulated by AGN activity rather than environmental conditions.

\subsection{Gas Fraction in AGN and non-AGN Galaxies}\label{sec:gas_fraction_comparison} 
\begin{figure}
    \centering
    \includegraphics[width=0.95\linewidth]{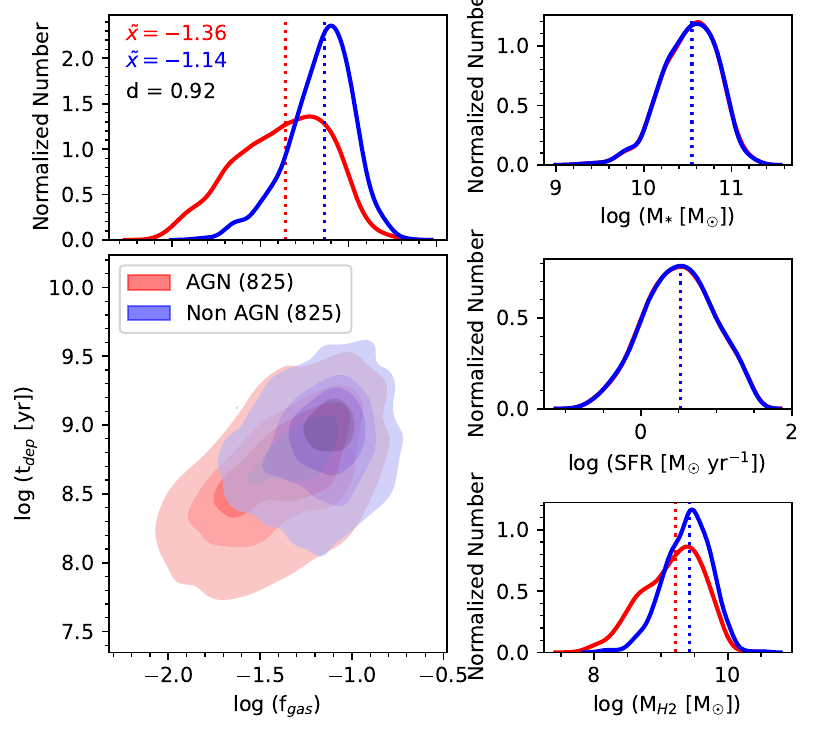}
    \caption{Comparison of different properties in type~2 AGN host (red) and non-AGN galaxies (blue), matched in stellar mass and SFR. }
    \label{fig:Match_AGN_SFG}
\end{figure}

Molecular gas plays a fundamental role in galaxy evolution, acting as the primary reservoir that fuels star formation and drives the growth of supermassive black holes. We compared ${\rm f_{gas}}$ of and type~2 AGN and non-AGN host galaxies regardless of the environment (Figure~\ref{fig:Match_AGN_SFG}). The type~2 AGN and non-AGN samples were matched in stellar mass and SFR. The cross-matching process resulted in 825 type~2 AGN galaxies and 825 non-AGN galaxies. The equal sample size of galaxies are chosen to avoid biases caused by unequal numbers of galaxies. The Cohen's d value for comparing ${\rm f_{gas}}$ is 0.92, indicating a significant difference in ${\rm f_{gas}}$ between type~2 AGN and non-AGN galaxies. The non-AGN galaxies have a higher ${\rm f_{gas}}$ relative to type~2 AGN host galaxies. \citet{Saintonge2017ApJS..233...22S} also demonstrated that, within the COLD GASS sample, galaxies matched in specific SFR show that BPT-selected AGN hosts possess slightly lower gas fractions compared to their inactive xCOLD GASS counterparts. This suggests that AGN activity may be linked to a depletion of the molecular gas reservoir, potentially due to the processes that either consume, expel, or heat the gas.

We also examined the ${\rm f_{gas}}$ of type~2 AGN and non-AGN galaxies using different SFR estimators, finding that the observed differences range from moderate to substantial. The ${\rm f_{gas}}$ of type~2 AGN and non-AGN galaxies, derived using matched SFR and ${\rm M_{*}}$ from \citet{Salim2018ApJ...859...11S} and \citet{Brinchmann2004MNRAS.351.1151B}, show modest differences. However, SFR estimates from \citet{Chang2015ApJS..219....8C} and \citet{Ayubinia2025ApJ...980..177A} reveals substantially larger ${\rm f_{gas}}$ differences between type~2 AGN and non-AGN galaxies. Thus, our results based on IR-derived SFR estimates are consistent with those obtained from other catalogs.

\subsubsection{Gas Fraction in Strong, Weak and No Outflow AGNs}
\begin{figure}
    \centering
    \includegraphics[width=0.95\linewidth]{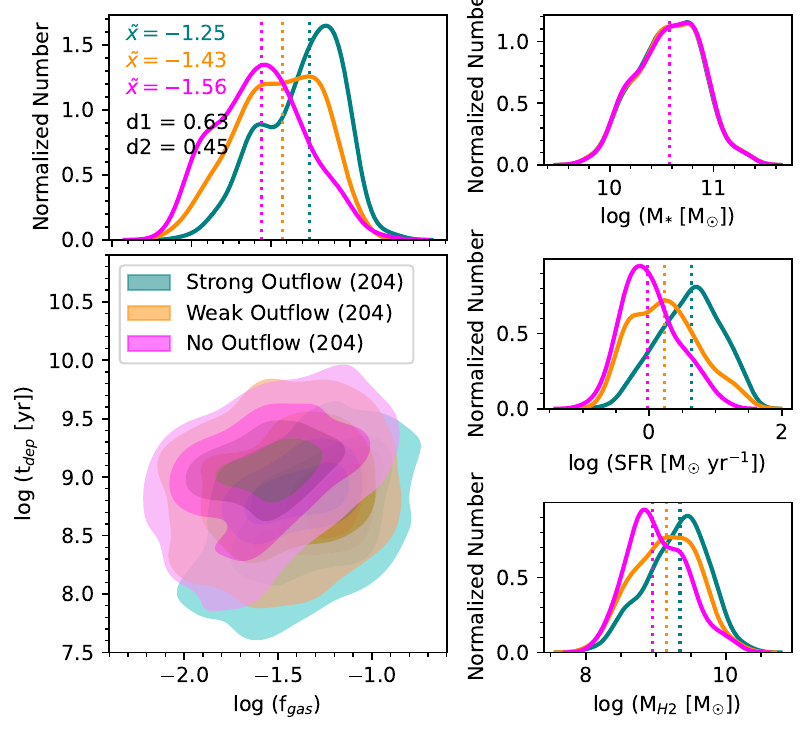}
    \caption{Comparison of properties of strong (teal), weak (orange), and no outflow AGNs (magenta) respectively. The dashed line shows the median value of each distribution. }
    \label{fig:AGN_SWN_gas}
\end{figure}

Understanding the role of gas content in driving AGN activity and feedback is essential for connecting galaxy evolution to its environment. To examine the correlation between ${\rm f_{gas}}$ and AGN outflows, we compared strong, weak, and no outflow type~2 AGNs matched in stellar mass (Figure~\ref{fig:AGN_SWN_gas}). AGNs with strong outflows exhibit higher SFR and sSFR than those with weaker or no outflows, with Cohen’s d values of 0.63 (strong vs. weak) and 0.45 (weak vs. none). ${\rm f_{gas}}$ follows the same trend, highest in strong outflow AGNs and progressively lower in weak and no outflow systems.

This pattern can be interpreted in two ways. First, galaxies rich in gas naturally sustain both high star formation and powerful AGN outflows, whereas depleted systems exhibit weaker outflows and reduced SFR, implying a direct link between gas availability and outflow strength. Second, AGN feedback may gradually deplete gas: initially abundant gas fuels strong outflows and star formation, but depletion occurs over time. Strong outflow AGNs thus represent an early or active evolutionary phase with substantial gas reservoirs, while weak and no outflow AGNs correspond to later stages, where gas has been suppressed or exhausted. This delayed feedback scenario aligns with \citet{Woo2017ApJ...839..120W}.

\begin{figure}
    \centering
    \includegraphics[width=0.95\linewidth]{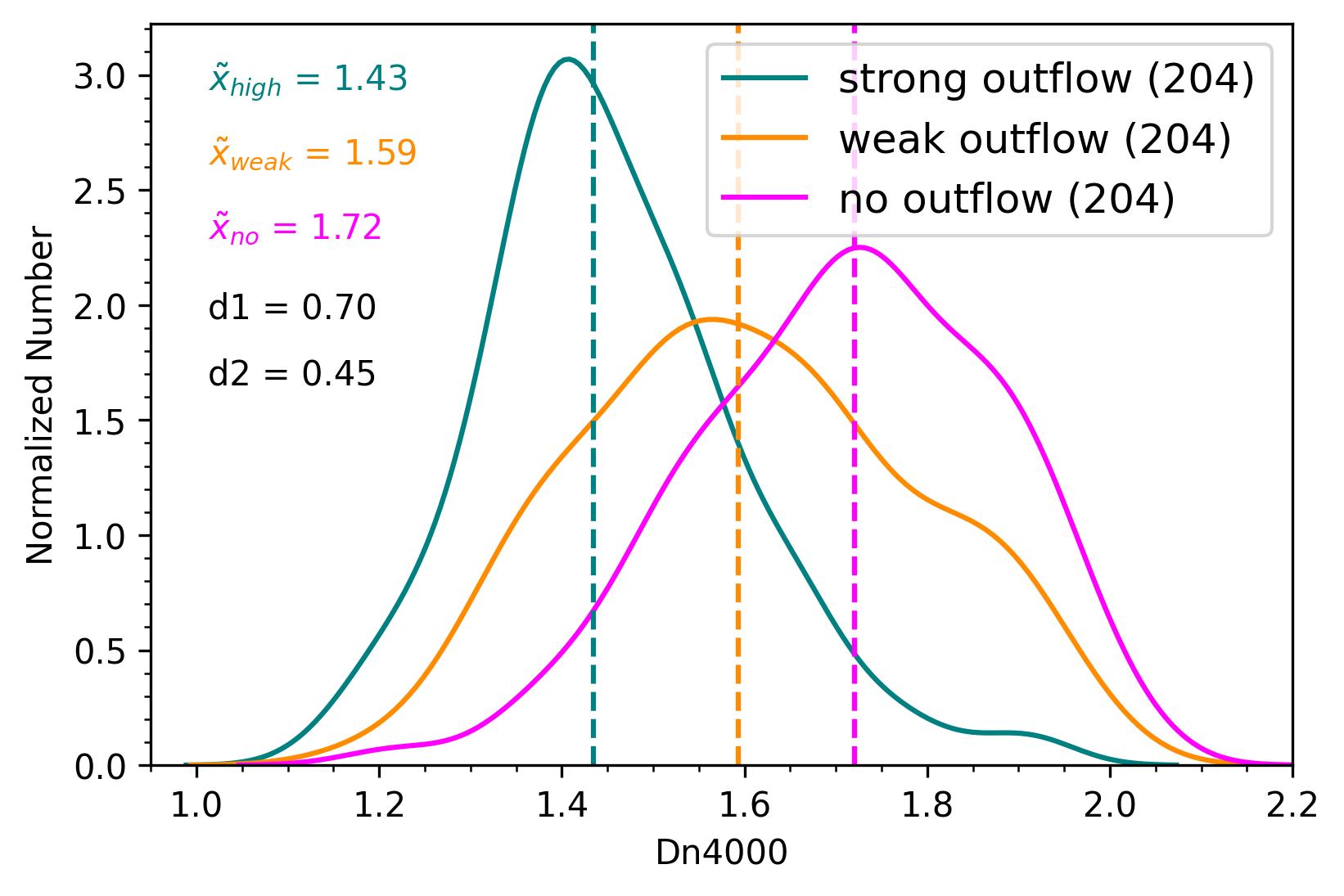} 
    \caption{${\rm D_{n}}$4000 in strong (teal), weak (orange), and no outflow AGNs (magenta) respectively. The dashed line shows the median value of each distribution.}
    \label{fig:Age_outflow}
\end{figure}

We examined D$_n$4000 as a stellar age indicator (Figure~\ref{fig:Age_outflow}). As expected, strong outflow AGNs with higher sSFR have lower D$_n$4000, reflecting younger stellar populations. This anti-correlation between D$_n$4000 and sSFR \citep{Kauffmann2003MNRAS.346.1055K} highlights that strong outflows occur in systems that are both more star-forming and younger.

\begin{figure}
    \centering
    \includegraphics[width=0.95\linewidth]{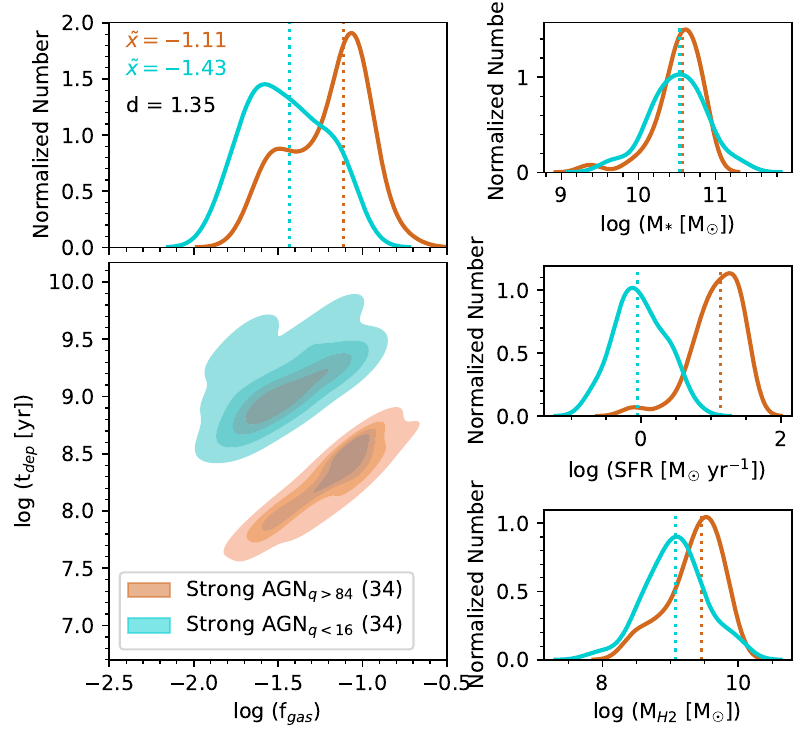} 
    \caption{Comparison of properties of above 84 percentile (brown) and below 16 percentile (cyan) sample of strong outflow AGNs. The dashed line shows the median value of each distribution.}
    \label{fig:percentile_strong_outflow}
\end{figure}

To test whether sSFR variations among strong outflow AGNs are driven by gas content, we compared ${\rm f_{gas}}$ for galaxies above the 84th and below the 16th percentile of sSFR (Figure~\ref{fig:percentile_strong_outflow}). High-sSFR systems have substantially higher ${\rm f_{gas}}$ than low-sSFR systems, with Cohen’s d = 1.35. This indicates that stronger outflows and higher sSFRs correspond to earlier or more active evolutionary stages, retaining significant gas for ongoing star formation and AGN activity, while lower-sSFR AGNs are likely more evolved with depleted reservoirs.

Observations and simulations indicate that black hole growth is variable and episodic \citep{Harrison2024Galax..12...17H}. AGNs can flicker multiple times, and outflows propagate through the host over time. For significant impact on global SFR, AGNs must experience recurrent activity before the gas reservoir fully recovers. In our study, AGNs currently lacking detectable outflows may represent systems that have already undergone multiple episodic events, leaving depleted gas reservoirs. In contrast, strong outflow AGNs retain high gas fractions and elevated sSFR, suggesting recent activity, whereas weaker outflows reflect older, more evolved phases of AGN feedback.

\section{Discussion} \label{sec:discussion}
\subsection{Environment Effect}
In this work, we examined the impact of large- and small-scale environments on type~2 AGN and non-AGN host galaxies. To probe large-scale effects, we evaluated the P, F, and T parameters, considering the cumulative influence of all galaxies within each group. Our results show no significant differences between type~2 AGN and non-AGN galaxies (Figure~\ref{fig:AGN_SFG_pft}), indicating that both populations experience similar gravitational effects from their environment. Furthermore, type~2 AGNs classified by outflow strength (strong, weak, or none) exhibit comparable P, F, and T distributions, suggesting that AGN-driven outflow strength is not linked to large-scale environmental factors. The type~2 AGN fraction also remains largely independent of these parameters.

Previous studies have reported a lower AGN fraction in galaxy clusters, attributed either to the reduced frequency of galaxy--galaxy interactions in dense environments \citep{Gisler1978MNRAS.183..633G, Dressler1985ApJ...288..481D, Popesso2006A&A...460L..23P, Pimbblet2013MNRAS.429.1827P, Lopes2017MNRAS.472..409L} or to the limited availability of cold gas for SMBH accretion \citep{Sabater2013MNRAS.430..638S}. Conversely, other works find that AGN fractions remain nearly constant across local densities \citep{Miller2003ApJ...597..142M, Martini2006ApJ...644..116M, Rodriguez2017MNRAS.467.4200R}, consistent with theoretical predictions that SMBH accretion is broadly suppressed in clusters across all mass ranges \citep{Joshi2020MNRAS.496.2673J}. Our results also shows no variation in AGN fraction with P, F, and T.

Examining the influence of galaxy environment on molecular gas, we find that large-scale factors, including group versus single galaxy membership, do not significantly affect ${\rm f_{gas}}$ or AGN activity (Figure~\ref{fig:AGN_SFG_group_single}). While some studies report enhanced molecular gas fractions and shorter depletion times in galaxy pairs \citep{Ellison2018MNRAS.478.3447E, Violino2018MNRAS.476.2591V, Yu2024ApJS..273....2Y, Sargent2025OJAp....8E..33S}, others find little correlation between environment and gas content at low \citep{Kenney1989ApJ...344..171K, Lavezzi1998AJ....115..405L, Koyama2017ApJ...847..137K} or high redshift \citep{Lee2017ApJ...842...55L, Tadaki2019PASJ...71...40T}, suggesting that molecular gas, being centrally concentrated and gravitationally bound, is largely resilient to environmental effects. Additionally, averaging the effects of all group members may homogenize environmental parameters, leading to consistent estimates across galaxies.

Analysis of nearest-neighbor effects further indicates that type~2 AGN hosts and non-AGN galaxies experience comparable small-scale environmental influences. However, when matched in stellar mass and environment, AGN hosts consistently exhibit lower ${\rm f_{gas}}$ than non-AGN counterparts. Given that the principal difference between the two samples is the presence of AGN activity, this result suggests that AGNs play a substantial role in depleting molecular gas reservoirs and consequently suppressing star formation.

\subsection{Gas Fraction}
\subsubsection{AGN Effect on Gas}
Differences in molecular gas properties between AGN and non-AGN galaxies provide key insights into the processes governing galaxy evolution. In this study, we compared ${\rm f_{gas}}$ of type~2 AGN hosts and non-AGN galaxies across a range of environments. Our estimates of ${\rm f_{gas}}$, derived from the NUV-r color, do not account for dust obscuration and may underestimate true gas fractions. While molecular gas correlates more strongly with sSFR, deriving ${\rm f_{gas}}$ from SFR--${\rm M_{*}}$ relations would artificially enforce similar distributions between AGN and non-AGN samples. Thus by using NUV-r colors, we reduces this bias while preserving meaningful differences between the populations \citep{Saintonge2017ApJS..233...22S, Salvestrini2025A&A...695A..23S}. However, note that ${\rm f_{gas}}$ is indirectly linked with sSFR as 
NUV-r color has been shown to correlate with sSFR \citep{Saintonge2011MNRAS.415...32S}. 

The influence of AGN on molecular gas remains debated. In the local Universe, AGN hosts are often indistinguishable from non-active galaxies in global gas content \citep{Rosario2018MNRAS.473.5658R, Koss2021ApJS..252...29K, Salvestrini2022A&A...663A..28S}, and the most powerful AGN can reside in gas-rich systems \citep{Vito2014MNRAS.441.1059V, Husemann2017MNRAS.470.1570H}. Nevertheless, negative feedback signatures, including molecular outflows, are observed \citep{Fiore2017A&A...601A.143F}, often confined to central regions \citep{Rosario2019ApJ...875L...8R, Fluetsch2019MNRAS.483.4586F, Feruglio2020ApJ...890...29F, Zanchettin2021A&A...655A..25Z, Garcia2021A&A...652A..98G, Garcia2024A&A...689A.347G, RamosAlmeida2023A&A...669L...5R}, suppressing star formation locally while leaving galaxy-wide SFR largely unaffected \citep{Sanchez2018RMxAA..54..217S, Lammers2023ApJ...953...26L}. \citet{Mountrichas2023A&A...675A.137M} found that isolated AGNs exhibit lower SFRs than non-AGNs across all X-ray luminosities. In denser environments (such as filaments and clusters), moderate-luminosity AGNs (log[${\rm L_{X,2-10\,keV} (erg\,s^{-1}})$] $>$ 43) show SFRs comparable to those of non-AGN galaxies. Furthermore, \citet{Mountrichas2024A&A...686A.229M} reported that the most luminous AGNs (log[${\rm L_{{X,2-10\,keV}} (erg\,s^{\rm -1})}$] $>$ 44) exhibit enhanced SFRs relative to non-AGNs, a trend that appears to be independent of environmental density. Some studies report enhanced central star formation or elevated star formation efficiencies in AGN hosts with large gas reservoirs \citep{Molina2023ApJ...944...30M, Bessiere2022MNRAS.512L..54B, Shangguan2020ApJS..247...15S, Bischetti2021A&A...645A..33B}. AGN variability further complicates direct comparisons \citep{Harrison2017NatAs...1E.165H, Harrison2024Galax..12...17H}. Recent work, however, shows CO- and HI-deficient AGN hosts relative to matched non-active galaxies \citep{Bertola2024A&A...691A.178B, Ellison2021MNRAS.505L..46E, Ellison2019MNRAS.482.5694E}, supporting a role for AGN in depleting the molecular gas reservoir.

We find that at fixed stellar mass and SFR, type~2 AGN hosts exhibit systematically lower ${\rm f_{gas}}$ than non-AGN galaxies (Figure~\ref{fig:Match_AGN_SFG}). This trend suggests that AGN activity regulates the cold gas reservoir available for star formation, likely through feedback mechanisms such as winds, jets, or radiative heating that expel or ionize the ISM. Alternatively, it may reflect a higher star formation efficiency, whereby the available molecular gas is consumed more rapidly. The progressive decline in ${\rm f_{gas}}$ from AGNs with strong outflows to those lacking clear outflow signatures underscores the cumulative influence of AGN feedback in suppressing star formation. Such processes are key to driving the transition of galaxies from star-forming to quiescent states \citep{Silk1998A&A...331L...1S, DiMatteo2005Natur.433..604D, Dekel2006MNRAS.368....2D, Martig2009ApJ...707..250M, Peng2015Natur.521..192P, Pillepich2018MNRAS.473.4077P}.

\subsubsection{Strong Outflow AGNs in Gas-Rich Galaxies}
The availability of molecular gas is a key regulator of both star formation and AGN activity. We find that type~2 AGNs with strong outflows exhibit higher ${\rm f_{gas}}$, elevated SFR, and younger stellar populations compared to weak or non-outflow AGNs (Figure~\ref{fig:Match_AGN_SFG}). This indicates that AGN feedback operates on dynamical timescales, rather than instantaneously affecting the molecular gas in  galaxies, consistent with observations. \citep{Netzer2009MNRAS.399.1907N, Diamond2012ApJ...746..168D, Esquej2014ApJ...780...86E, Delvecchio2015MNRAS.449..373D, Shangguan2020ApJ...899..112S, Yesuf2020ApJ...901...42Y} and theoretical models \citep{Hopkins2010MNRAS.407.1529H, Thacker2014MNRAS.443.1125T, Volonteri2015MNRAS.449.1470V, McAlpine2017MNRAS.468.3395M} suggesting that strong AGNs coexist with gas-rich star-forming hosts.

A delayed feedback scenario \citep{Woo2020ApJ...901...66W} is supported by both simulations, showing that AGN-driven jets can heat halo and circumgalactic gas over extended timescales \citep{Ciotti2001ApJ...551..131C, Ciotti2007ApJ...665.1038C, McNamara2007ARA&A..45..117M}, and observations of outflows inducing gradual suppression of star formation \citep{Sanchez2018RMxAA..54..217S}. In this framework, strong outflow AGNs have not yet had time to deplete their gas, whereas weak and non-outflow AGNs, likely experiencing longer AGN episodes, exhibit lower ${\rm f_{gas}}$, reduced SFR, and older stellar populations, consistent with progressive gas depletion and quenching.

Environmental factors further modulate star formation. Among strong outflow AGNs, galaxies with higher sSFR show lower P, F, and T values and higher ${\rm f_{gas}}$ than those with lower sSFR, suggesting that local and large-scale environment, together with gas content, influence star formation. Processes such as ram-pressure stripping \citep{Gunn1972ApJ...176....1G, Hester2006ApJ...647..910H, Brown2017MNRAS.466.1275B, Singh2019MNRAS.489.5582S}, strangulation \citep{Larson1980ApJ...237..692L}, and gravitational interactions \citep{Moore1998ApJ...495..139M, Moore1999MNRAS.304..465M} can all affect gas reservoirs. Given the limited sample size, larger studies are required to robustly confirm these trends and disentangle the interplay between AGN feedback and environment.

\section{Summary and Conclusions} \label{sec:summary}
In this study, we investigated the environmental effects and ${\rm f_{gas}}$ in type~2 AGN and non-AGN galaxies using a large sample of galaxies from SDSS DR7 up to z $\le$ 0.3.
\begin{enumerate}
    \item We find that the distributions of environmental parameters P, F, and T are similar for type~2 AGN and non-AGN galaxies. This suggests that large-scale environmental factors influence type~2 AGN hosts and non-AGN galaxies in similar ways. 
    \item We find that the distributions of the P, F, and T parameters for strong, weak, and no outflow type~2 AGNs are similar. This result suggests that environmental factors do not significantly influence the outflow strength in AGNs.
    \item The SFRs and ${\rm f_{gas}}$ of type~2 AGN host galaxies show no significant differences between group and isolated environments. This indicates that internal processes predominantly govern the properties of galaxies, rather than large-scale environmental effects.
    \item We do not find a significant difference in the SFRs and ${\rm f_{gas}}$ of the group and isolated non-AGN galaxies. This suggests that similar to AGN host galaxies, internal processes are more influential than large-scale environmental factors in determining the properties of these galaxies.
    \item We find that type~2 AGNs with strong outflows exhibit higher ${\rm f_{gas}}$ compared to those with weak or no outflows. 
    \item We find that non-AGN galaxies exhibit systematically higher ${\rm f_{gas}}$ than type~2 AGN host galaxies, independent of environment, both in group and isolated systems. This difference may be linked to AGN feedback processes, which can regulate the gas content of AGN hosts through mechanisms such as heating or expulsion of the interstellar medium, or through more rapid gas consumption associated with enhanced star formation efficiency. In contrast, non-AGN galaxies, which are not subject to strong AGN feedback, are able to retain larger cold gas reservoirs and therefore display higher ${\rm f_{gas}}$.
\end{enumerate}

In this study, we have utilized galaxy color as a proxy for estimating ${\rm M_{H_{2}}}$, allowing us to explore trends in ${\rm f_{gas}}$ and depletion time across different populations of galaxies, including type~2 AGN hosts and non-AGN galaxies. While this approach provides valuable insights into the broad relationships between AGN activity and the gas reservoirs of galaxies, it remains an indirect method that may introduce uncertainties due to the complex interplay of factors influencing galaxy colors, such as dust attenuation, star formation history, and metallicity. Direct observations of molecular gas, such as those enabled by CO emission line surveys or dust continuum measurements for a larger sample of AGN host galaxies, will provide a more accurate and detailed understanding of the gas content in these systems. These observations will also help to clarify the spatial distribution of gas, the role of AGN feedback in shaping gas reservoirs, and the mechanisms governing gas depletion. By combining high-resolution molecular gas observations with other tracers, future studies can more precisely quantify the relationship between ${\rm f_{gas}}$, depletion time, and AGN activity, ultimately shedding light on the processes that regulate galaxy growth and evolution.

\begin{acknowledgments}
We thank the anonymous referee for a careful reading of the manuscript and for constructive comments and suggestions that significantly improved the clarity and quality of this work. This work has been supported by the Basic Science Research Program through the National Research Foundation of Korean Government (2021R1A2C3008486). JY acknowledges financial support from the  Spanish Ministry of Science, Innovation and Universities (Ministerio de Ciencia, Innovación y Universidades, MICIU) , project PID2022-136598NBC31 (ESTALLIDOS8) by
MCIN/AEI/10.13039/501100011033.
\end{acknowledgments}

\appendix
\section{P F T Parameters}\label{sec:pft_analysis}
The P, F, and T parameters offer valuable insights into the environmental influence on galaxies within a group. Specifically, P quantifies the total gravitational potential, representing the cumulative effect of all other group members, F captures the sum of force magnitudes, and T reflects the total tidal field strength experienced by a galaxy due to its neighbors.

\begin{figure*}
\centering
    \includegraphics[width=0.95\linewidth]{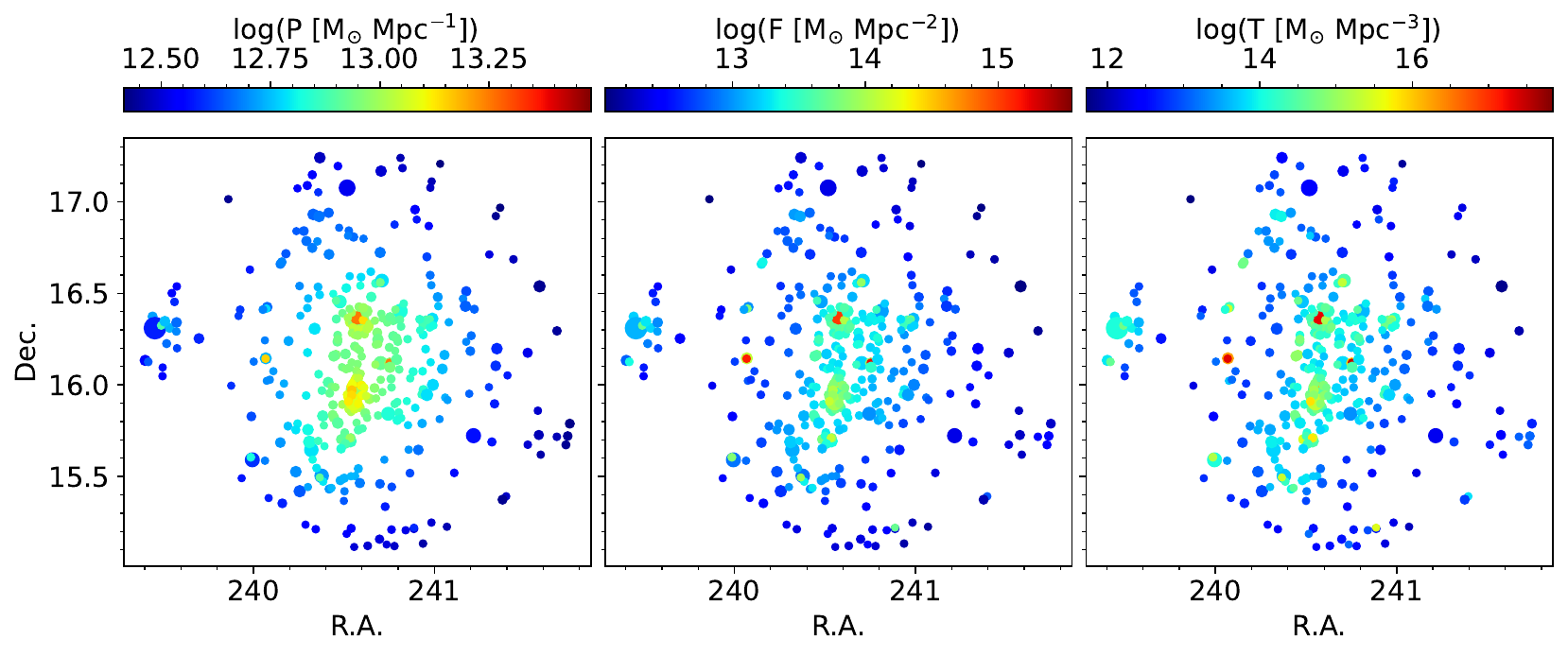}
    \caption{The spatial distribution of galaxies within a group, color-coded by their P, F, and T parameters. The galaxy sizes are scaled according to their masses.}
    \label{fig:demo_pft}
\end{figure*}

Figure~\ref{fig:demo_pft} shows the spatial distribution of these parameters within a galaxy group. The galaxies in the central regions exhibit higher P parameter (left panel) values in the central region and the outskirts galaxies show lower P values. This trend is primarily driven by the presence of two massive galaxies at the center, whose gravitational influence dominates the P values of surrounding galaxies. However, due to its inverse dependence on distance r, this influence weakens for galaxies located further out.

\begin{figure*}
    \centering
    \includegraphics[width=0.95\linewidth]{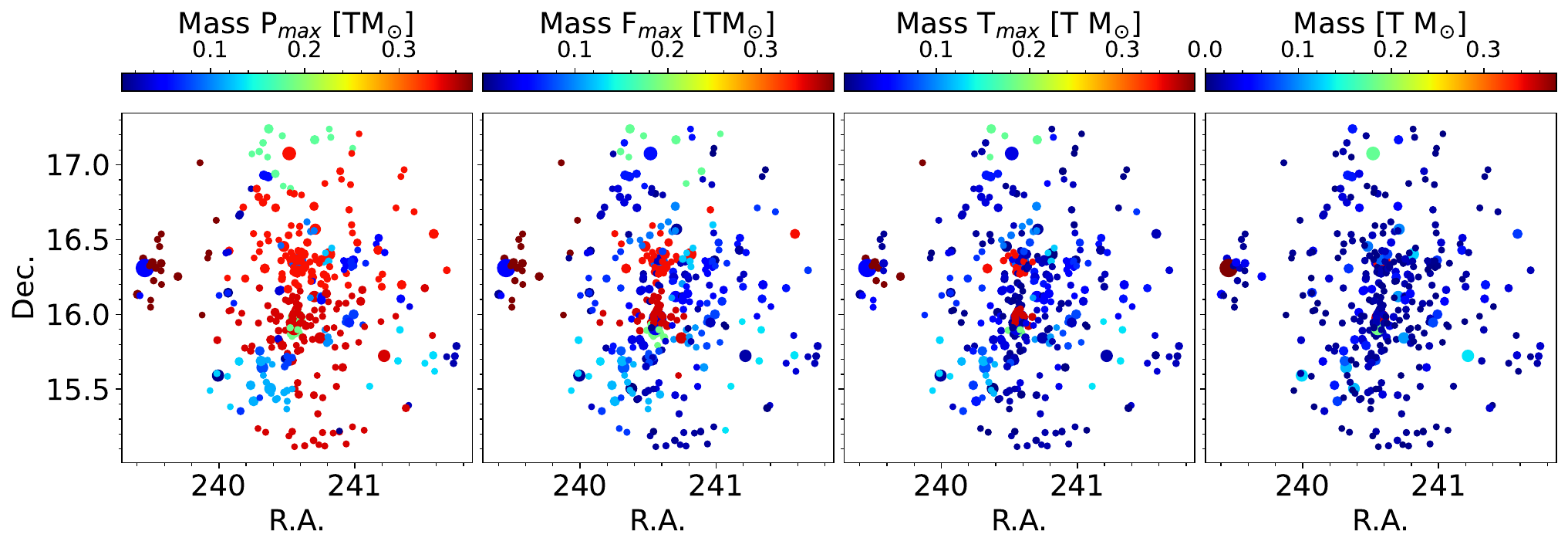}
    \caption{The spatial distribution of galaxies in a group. The first three plots show galaxies color-coded by the mass of the dominant contributor to their P, F, and T parameters. The rightmost plot represents galaxies colored and sized by their own masses.}
    \label{fig:demo_pftmax}
\end{figure*}

The F parameter, which scales with 1/r$^{2}$, becomes more localized in comparison to P. As a result, only nearby galaxies significantly contribute to F, highlighting more immediate gravitational interactions. The T parameter, with an even stronger localization, is dominated by nearby massive galaxies alone. Interestingly, the central galaxies do not always play a major role in the T values of outer group members. Thus, while P reflects the broader-scale environmental potential, T is more sensitive to the local-scale structure.  Figure~\ref{fig:demo_pftmax} demonstrates this by mapping the spatial distribution of galaxies, color-coded according to the mass of the dominant contributing galaxy for each parameter. The P parameter (left panel) is influenced by the two central massive galaxies across much of the group. However, in the F panel, their impact becomes confined to nearby galaxies, illustrating the increasingly localized nature of environmental influence. The T panel shows an even sharper transition, with only the closest neighbors significantly affected by the central or other massive galaxies. 
The mass distribution of the group (right panel) shows multiple massive galaxies. Their localized environmental effects are most evident in the T parameter map, underscoring how gravitational interactions shift from global to local scales as we move from P to T.

The P, F, and T parameters together provide a multiscale perspective on environmental effects: P captures the large-scale gravitational effect, while T highlights the finer, more immediate interactions.

\section{Comparisons between different SFR estimates}\label{sec:comparison_sfr}
Several SFR estimates are available in the literature, each using different tracers and methodologies. The MPA-JHU catalog provides SFRs based on galaxy classification \citep{Brinchmann2004MNRAS.351.1151B}. For star-forming galaxies, SFRs are derived from dust-corrected H$\alpha$ emission, with extinction estimated using the Balmer decrement. For AGNs and weak emission line galaxies, SFRs are estimated from the 4000$\AA$ break strength following the method of \citet{Kauffmann2003MNRAS.346.1055K}. All SFRs are corrected for fiber aperture effects and re-calibrated to the Chabrier IMF \citep{Chabrier2003PASP..115..763C}.

The GSWLC-2 catalog provides SFRs from UV/optical SED fitting and mid-infrared measurements based on 22 $\mu$m WISE data \citep{Salim2018ApJ...859...11S}.

\citet{Chang2015ApJS..219....8C} presented SFRs derived from combined SDSS and WISE photometry, constructing SEDs covering 0.4--22$\mu$m for over 850,000 local galaxies. These SFRs were primarily based on PAH emission and thermal dust radiation and include updated calibrations for converting monochromatic 12 and 22$\mu$m luminosities into SFRs.

\citet{Ayubinia2025ApJ...980..177A} estimated SFRs by predicting infrared luminosities using an artificial neural network trained on true IR luminosities obtained from detailed modeling of FIR-detected galaxy SEDs.

\begin{figure*}
    \centering
    \includegraphics[width=0.43\linewidth]{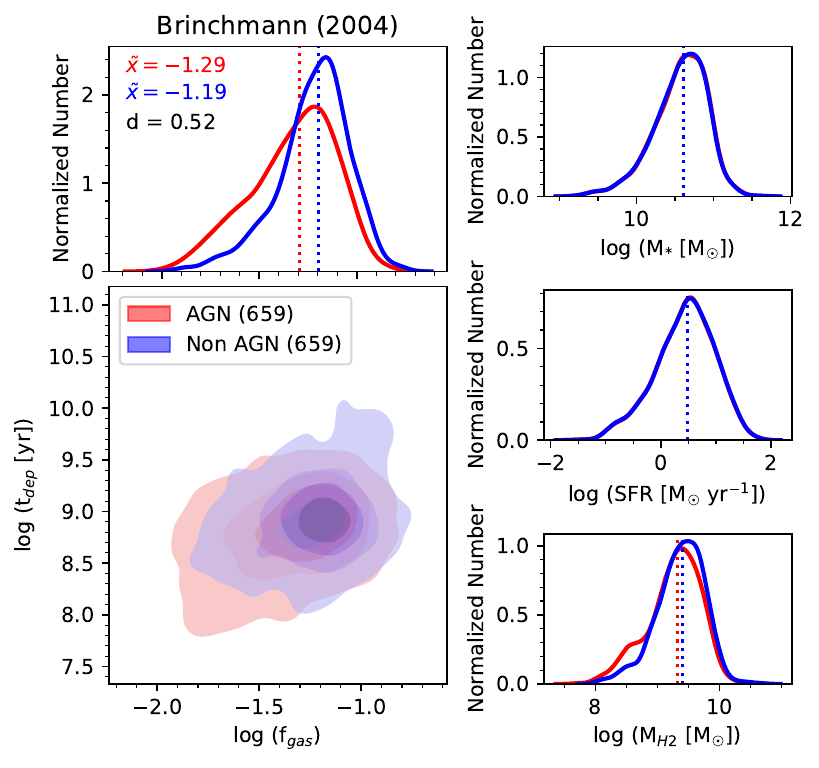} 
    \includegraphics[width=0.43\linewidth]{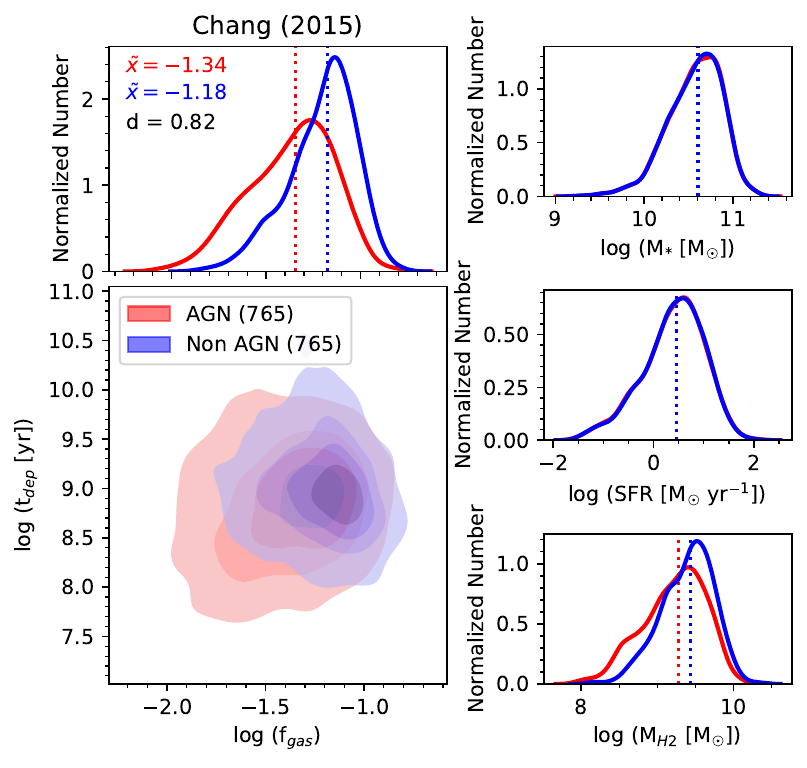}
    \includegraphics[width=0.43\linewidth]{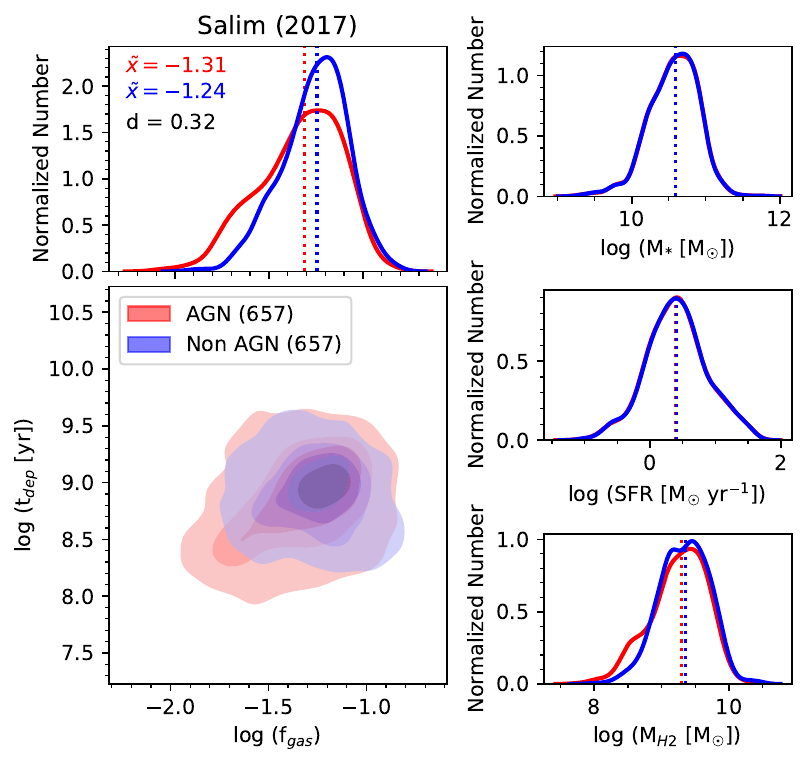}
    \includegraphics[width=0.43\linewidth]{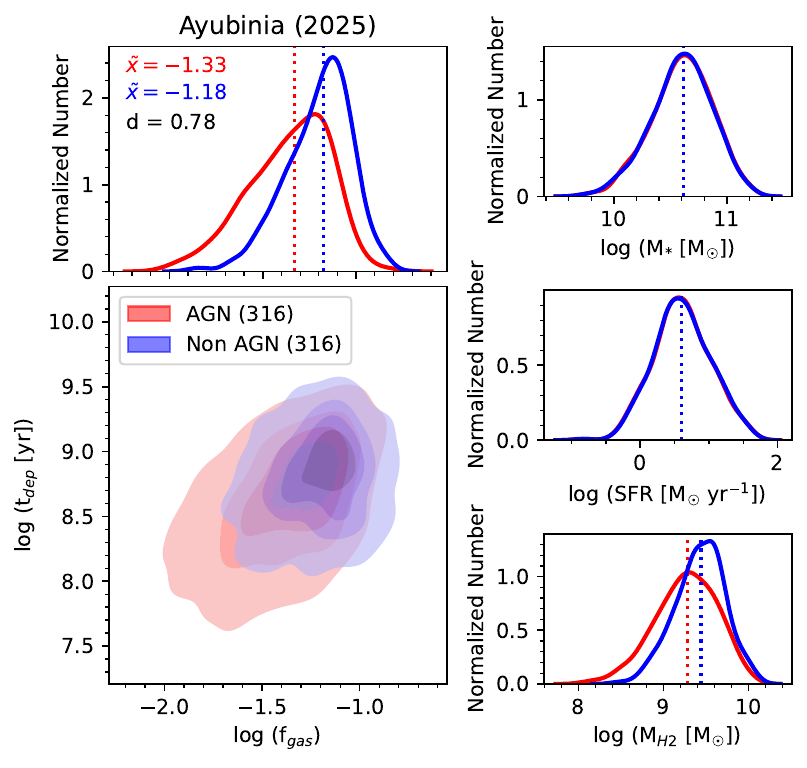}
    \caption{Comparison of the properties of AGN and non-AGN galaxies using different SFR estimates.}
    \label{fig:enter-label}
\end{figure*}




\bibliography{references}{}
\bibliographystyle{aasjournal}



\end{document}